\documentclass[twoside,twocolumn,9pt]{article}
\usepackage{extsizes}
\usepackage[super,sort&compress,comma]{natbib}
\usepackage[version=3]{mhchem}
\usepackage[left=1.5cm, right=1.5cm, top=1.785cm, bottom=2.0cm]{geometry}
\usepackage{balance}
\usepackage{times,mathptmx}
\usepackage{sectsty}
\usepackage{graphicx}
\usepackage{lastpage}
\usepackage[format=plain,justification=justified,singlelinecheck=false,font={stretch=1.125,small,sf},labelfont=bf,labelsep=space]{caption}
\usepackage{float}
\usepackage{fancyhdr}
\usepackage{fnpos}
\usepackage[english]{babel}
\addto{\captionsenglish}{%
  \renewcommand{\refname}{Notes and references}
}
\usepackage{array}
\usepackage{droidsans}
\usepackage{charter}
\usepackage[T1]{fontenc}
\usepackage[usenames,dvipsnames]{xcolor}
\usepackage{setspace}
\usepackage[compact]{titlesec}
\usepackage{hyperref}

\definecolor{cream}{RGB}{222,217,201}

\usepackage{amsmath, mathtools} 
\usepackage[utf8]{inputenc}
\usepackage{subcaption}
\usepackage{amssymb}

\usepackage{physics}

\hypersetup{colorlinks=true, linkcolor=NavyBlue, citecolor=PineGreen,urlcolor=cyan}
\usepackage{tikz}

\newcommand{\avg}[1]{\left< #1 \right>} 

\allowdisplaybreaks
\numberwithin{equation}{section}

\graphicspath{{./Figures/}}

\begin{document}

\pagestyle{fancy}
\thispagestyle{plain}
\fancypagestyle{plain}{}


\makeFNbottom
\makeatletter
\renewcommand\LARGE{\@setfontsize\LARGE{15pt}{17}}
\renewcommand\Large{\@setfontsize\Large{12pt}{14}}
\renewcommand\large{\@setfontsize\large{10pt}{12}}
\renewcommand\footnotesize{\@setfontsize\footnotesize{7pt}{10}}
\makeatother

\renewcommand{\thefootnote}{\fnsymbol{footnote}}
\renewcommand\footnoterule{\vspace*{1pt}%
\color{cream}\hrule width 3.5in height 0.4pt \color{black}\vspace*{5pt}}
\setcounter{secnumdepth}{5}

\makeatletter
\renewcommand\@biblabel[1]{#1}
\renewcommand\@makefntext[1]%
{\noindent\makebox[0pt][r]{\@thefnmark\,}#1}
\makeatother
\renewcommand{\figurename}{\small{Fig.}~}
\sectionfont{\sffamily\Large}
\subsectionfont{\normalsize}
\subsubsectionfont{\bf}
\setstretch{1.125} 
\setlength{\skip\footins}{0.8cm}
\setlength{\footnotesep}{0.25cm}
\setlength{\jot}{10pt}
\titlespacing*{\section}{0pt}{4pt}{4pt}
\titlespacing*{\subsection}{0pt}{15pt}{1pt}

\fancyfoot{}
\fancyfoot[LO,RE]{\vspace{-7.1pt}\includegraphics[height=9pt]{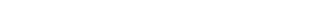}}
\fancyfoot[CO]{\vspace{-7.1pt}\hspace{13.2cm}\includegraphics{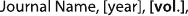}}
\fancyfoot[CE]{\vspace{-7.2pt}\hspace{-14.2cm}\includegraphics{head_foot/RF}}
\fancyfoot[RO]{\footnotesize{\sffamily{1--\pageref{LastPage} ~\textbar  \hspace{2pt}\thepage}}}
\fancyfoot[LE]{\footnotesize{\sffamily{\thepage~\textbar\hspace{3.45cm} 1--\pageref{LastPage}}}}
\fancyhead{}
\renewcommand{\headrulewidth}{0pt}
\renewcommand{\footrulewidth}{0pt}
\setlength{\arrayrulewidth}{1pt}
\setlength{\columnsep}{6.5mm}
\setlength\bibsep{1pt}

\makeatletter
\newlength{\figrulesep}
\setlength{\figrulesep}{0.5\textfloatsep}

\newcommand{\topfigrule}{\vspace*{-1pt}%
\noindent{\color{cream}\rule[-\figrulesep]{\columnwidth}{1.5pt}} }

\newcommand{\botfigrule}{\vspace*{-2pt}%
\noindent{\color{cream}\rule[\figrulesep]{\columnwidth}{1.5pt}} }

\newcommand{\dblfigrule}{\vspace*{-1pt}%
\noindent{\color{cream}\rule[-\figrulesep]{\textwidth}{1.5pt}} }

\makeatother

\twocolumn[
  \begin{@twocolumnfalse}
	{\includegraphics[height=30pt]{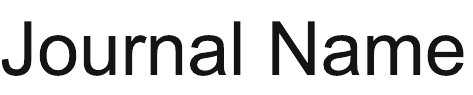}\hfill%
	 \raisebox{0pt}[0pt][0pt]{\includegraphics[height=55pt]{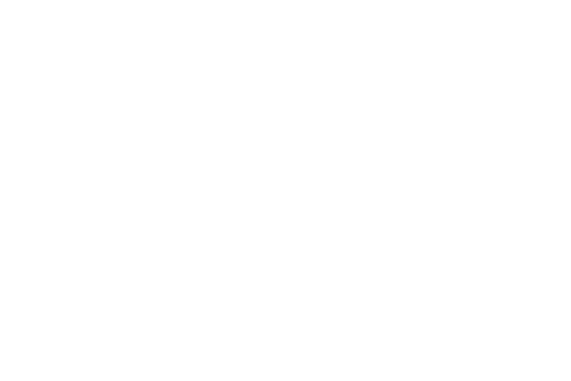}}%
	 \\[1ex]%
	 \includegraphics[width=18.5cm]{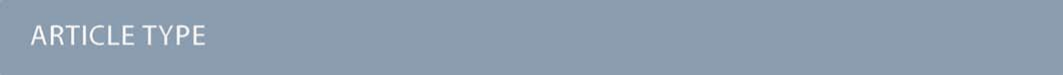}}\par
%
%
%
\vspace{1em} 
\sffamily
\begin{tabular}{m{4.5cm} p{13.5cm} }

\includegraphics{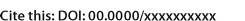} & \noindent\LARGE{\textbf{Inferring the particle-wise dynamics of amorphous solids from the local structure at the jamming point}} \\
\vspace{0.3cm} & \vspace{0.3cm} \\

 & \noindent\large{Rafael Díaz Hernández Rojas,$^{\ast}$\textit{$^{a}$} Giorgio Parisi,\textit{$^{a,b,c}$} and Federico Ricci-Tersenghi\textit{$^{a,b,c}$}} \\

\includegraphics{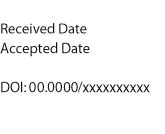} & \noindent\normalsize{Jamming is a phenomenon shared by a wide variety of systems, such as granular materials, foams, and glasses in their high density regime. This has motivated the development of a theoretical framework capable of explaining many of their static critical properties with a unified approach. However the dynamics occurring in the vicinity of the jamming point has received little attention and the problem of finding a connection with the local structure of the configuration remains unexplored. Here we address this issue by constructing physically well defined structural variables using the information contained in the network of contacts of jammed configurations, and then showing that such variables yield a resilient statistical description of the particle-wise dynamics near this critical point.
Our results are based on extensive numerical simulations of systems of spherical particles that allow us to statistically characterize the trajectories of individual particles in terms of their first two moments. We first demonstrate that, besides displaying a broad distribution of mobilities, particles may also have preferential directions of motion. Next, we associate each of these features with a structural variable computed uniquely in terms of the contact vectors at jamming, obtaining considerably high statistical correlations. The robustness of our approach is confirmed by testing two types of dynamical protocols, namely Molecular Dynamics and Monte Carlo, with different types of interaction.
We also provide evidence that the dynamical regime we study here is dominated by anharmonic effects and therefore it cannot be described properly in terms of vibrational modes.
Finally, we show that correlations decay slowly and in an interaction-independent fashion, suggesting a universal rate of information loss.
} \\

\end{tabular}

 \end{@twocolumnfalse} \vspace{0.6cm}

  ]

\renewcommand*\rmdefault{bch}\normalfont\upshape
\rmfamily
\section*{}
\vspace{-1cm}


\footnotetext{\textit{$^{a}$~Dipartimento di Fisica, Sapienza Università di Roma - Piazzale A. Moro 2, I-00185 Rome, Italy; E-mail: rafael.diazhernandezrojas@uniroma1.it}}
\footnotetext{\textit{$^{b}$~INFN-Sezione di Roma 1 - Piazzale A. Moro 2, I-00185 Rome, Italy }}

\footnotetext{\textit{$^{c}$~Nanotec-CNR, Rome unit, Sapienza Università di Roma - Piazzale A. Moro 2, I-00185, Rome, Italy}}




\section{Introduction}

Research on amorphous solids has been ever increasing for some time now, specially due to several striking properties these materials display when their density increases and their dynamics slows down. In addition to these phenomena, which by themselves have proved to be as interesting as they are hard to explain satisfactorily, another breakthrough arrived when a common theoretical background was conceived to tackle what at first seemed widely unrelated areas such as granular materials, foams and supercooled liquids well below their glass transition temperature\cite{torquato_review_jamming,liu_jamming_annrev,hecke_jamming_2010,siemens_jamming:_2010}. This unified viewpoint developed from the critical properties that all these systems posses near the so called ``jamming point'', which can be identified as a common critical point in their corresponding phase diagrams, and includes the respective physical variables needed to describe these systems macroscopically, \textit{e.g.} temperature and specific volume in supercooled liquids, or bulk modulus and packing fraction in amorphous solids. What is more, since the pioneering works of Nagel and Liu\cite{liu_nagel_nature,ohern_jamming_2003,liu_jamming_annrev} it became clear that the jamming point is not only a shared critical point for all these systems, but more importantly, it can also be used as a reference point to study their properties near it.
This promising picture fostered a huge theoretical effort to achieve a complete theory of jamming\cite{liu_jamming_annrev,parisi_zamponi-review}, and although this is still to be accomplished in the most general case, recently a comprehensive mean field theory of the jamming transition was obtained for the case of hard spheres in infinite dimensions\cite{puz_TheorySimpleGlasses2020,exact_theory-1,exact_theory-2,exact_theory-3,fractal-free-energy,parisi-et-al-jamming-annrev}.
Furthermore, many of their predictions were then proved to carry over all the way down to $2d$ and $3d$ systems\cite{jamming_criticality_forces,charbonneau_universal_2016,gardner_physics_crytals,gardner-perspective_2019,hagh_broader_2019} with outstanding accuracy. It should also be mentioned that these and other modern studies\cite{jiao_nonuniversality_2011} have shown that systems reach their jammed states not at a well defined point, but rather inside a rather broad region, so a more accurate term would be to refer to a ``jamming line'' or even ``jamming plane''. For brevity however, throughout this work we will maintain the more common phrasing and refer to \emph{the} jamming point, as if all the jammed states belong to it.
Similarly, for the sake of generality, and with a slight abuse of terminology, we will generically refer to all the systems mentioned as “glassy systems”.

On the other hand, as soon as one moves away from jamming, the degrees of freedom of a system are no longer completely blocked and particles begin to move in a rather complex manner, which renders a complete physical description even more complicated. For instance, since the discovery that glassy systems exhibit heterogeneous dynamics\cite{brito_heterogeneous_2007,berthier_theoretical_perspective,berthier_dynamical_2011,candelier_spatiotemporal_2010}, people have been looking for their connection with different structural, and hence \emph{static}, properties. Most of the works proceed in a similar fashion, namely by first calculating the mobility of individual particles (where this quantity is measured by the squared displacement; see Eq~\eqref{seq:square displacement} below) and then using the interactions coming from their nearby environment to construct physical observables with the expectation of finding a link between them. It is worth emphasizing that establishing this relation is by no means a trivial task, and in the search for significant correlation levels, several structural quantities have been proposed like soft spots\cite{soft_spots_manning}, local thermal energy\cite{local_thermal_energy}, point-to-set correlations\cite{berthier_static_2012,hocky_growing_2012,charbonneau_pts,karmakar_length_2016}, or even simple geometrical parameters like bond orientations\cite{tong_prx,tong_prl}, see \cite{review-local-structure} for a review.
A particularly fruitful approach, specially near the jamming regime, is based on the analysis of the vibrational modes of the system\cite{silbert_normal_2009, dynamic-criticality-jamming,wyart_effects_compression,wyart_geometric_2005,charbonneau_universal_2016,arceri_vibrational_2020,suggestion-1,suggestion-2}, and relating the density of states (DOS) with the collective dynamics of the system. This link is justified because the Fourier transform of the velocity autocorrelation function is closely related to the DOS\cite{pedagogical-vibrational-modes}. For this relation to be valid however, a harmonic approximation should hold for the interaction energy of the system, a scenario that in several cases might not be true.
Alternatively, a different and rather more intuitive approach to the same problem can be obtained by working within the so called “isoconfigurational ensemble”\cite{widmer-cooper_how_2004,widmer-cooper_relationship_2005,widmer-cooper_study_2007,jack_information-theoretic_2014} where the dynamics of a system in a given configuration are studied by performing several realizations of, say, molecular dynamics (MD) simulations that are initialised with the same particles' positions while the velocities are randomly assigned at each run according to the Maxwell-Boltzmann distribution with a fixed temperature. It is clear that this method provides a way of sampling the space of all possible trajectories of the configuration and in doing so allows to identify how the local environment of a particle influences its mobility since the contribution coming from the thermal noise is washed out when a large enough number of MD realizations are used. Specifically, regions where clusters of particles undergo major rearrangements during their dynamics can be identified and, more importantly, can also be related to structural properties such as the Debye--Waller factor or the local energy density\cite{jack_information-theoretic_2014,widmer-cooper_irreversible_2008,widmer-cooper_predicting_2006}. However, it has also been pointed out that using solely a particle's mobility to provide a link between dynamical and statical properties might be inadequate\cite{berthier_structure_2007}, while the results are highly sensitive to the type of glass former model used\cite{hocky_correlation_2014}.

Recently, a major step forward has been achieved along the same lines of finding a connection between the dynamics of particles and its local structure by employing Machine Learning methods\cite{schoenholz_identifying_2015,schoenholz_combining_2018}. In these works, the information of a particle's local structure is encoded in a new variable termed ``softness'' that is then related to how much this same particle has moved in a given time interval. A Support Vector Machine is used to find a hyperplane (in features space) dividing the mostly movable particles from the mostly arrested ones. Then, the softness is computed as the signed distance of each particle's features to such an hyperplane, thus ``soft'' particles are prone to be displaced by a significant amount while ``hard'' ones will remain mostly fixed. It has been shown that softness is strongly correlated with physical quantities such as local energy and coordination number\cite{schoenholz_structural_2016}, as well as being a useful variable for modelling the Arrhenius behaviour observed in supercooled liquids\cite{schoenholz_relationship_2017,schoenholz_combining_2018,schoenholz_nature_2016}. However, even though a Machine Learning inferential approach yields high quality predictions about which particles are the likeliest to move significantly, it also lacks a clear physical interpretation due to the synthetic representation of the particle's local environment as a consequence of its parametrization in terms of feature functions. In short, even if Machine Learning methods can be used to construct predictors of a particles mobility, they fail to provide an answer about which are the real physical variables that determine that mobility.

This ample variety of approaches, which often give results that are not consistent among them, is a signature of the intricacy of the problem of finding a correspondence between structural properties of glassy systems and their dynamics. In an attempt to get rid of several difficulties addressed by the studies mentioned above, in the present work we will make use of a simple glass former, namely, a monodisperse system composed of frictionless, spherical particles, and, more importantly, investigate the static-dynamic connection very near the jamming point. It should be stressed that this is a regime that has received little attention from a theoretical and numerical point of view, (Refs.~\cite{dynamic-criticality-jamming,procaccia-agitated-granular} are notable exceptions, albeit in a somewhat different scenario from the one we will consider here). This fact is even more surprising considering that experimental techniques allow to precisely probe this type of dynamics\cite{exper_dauchot_dynamical_2005,exper_lechenault_critical_2008,exper_brito_elementary_2010,exper_deseigne_collective_2010,exper_deseigne_vibrated_2012,exper_coulais_how_2014,experimental_gardner_olivier}. The choice for studying such a constrained dynamical regime is guided by the fact that the jamming point is a critical point where any motion of the configuration is prevented solely due to geometric frustration: \textit{e.g.} for spherical particles in any dimension, a jammed stated is uniquely determined by their position. Therefore we expect that under these circumstances any possible connection between structural properties of individual particles and their dynamics should be more noticeable. Our starting point will be the network of contact forces that is formed at jamming, whereby we will construct simple, yet robust, variables that allow us to characterize, individually, the displacement of particles. Specifically, our description identifies the most mobile particles but also if they exhibit any preferential direction of motion, a point that has been rarely addressed. In this way, we aim at dispelling many of the issues mentioned above since, on the one hand, the network of contact forces is a well defined physical quantity and the simplicity of the structural variables we consider -- see Eqs.~\eqref{def:tot CV} and \eqref{def:tot dot CVs} below-- makes our method considerably versatile, while on the other, we are of the mind that our approach provides a more detailed description of the particles motion, because we are able to identify any inherent anisotropy in the particles trajectory rather independently of the magnitude of their displacement.
Before concluding this section, we mention that our assumption that the statics has some bearing with the dynamics is motivated by the picture of the (fractal) free energy landscape of structural glasses\cite{fractal-free-energy}. Let us recall that within this framework a jammed system --where obviously there is no motion whatsoever-- can be thought of as being in one of the many possible minima of either a system of soft particles at temperature $T=0$ or one composed of infinitely hard ones at infinite pressure, while the dynamics takes place as the system explores the associated basin and possibly the neighbouring ones. Hence, it is reasonable to expect that if one of these minima is used as initial condition for the dynamics, the trajectory of the configuration as it moves in phase space should be influenced by the specific jammed stated from where it departed, at least for short times.

\begin{figure*}[h!]
	\centering
	\begin{subfigure}{0.4\textwidth}
  \includegraphics[width=\textwidth]{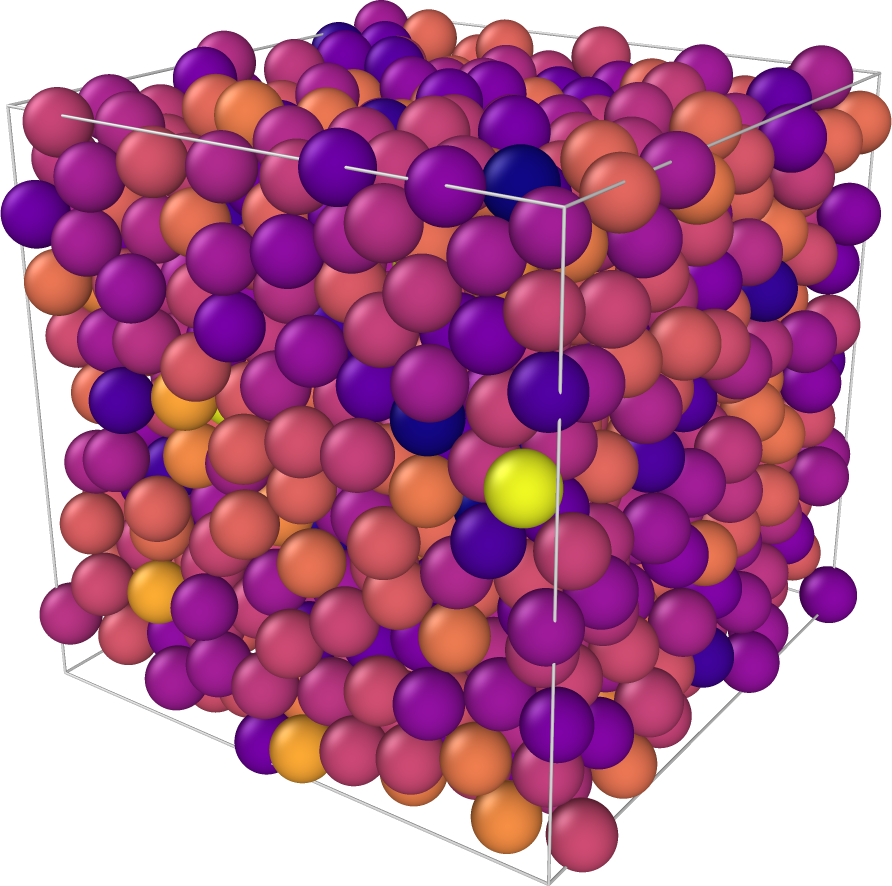}
	\end{subfigure} \hfil
	\begin{subfigure}{0.4\textwidth}
	\centering
	\includegraphics[width=\textwidth]{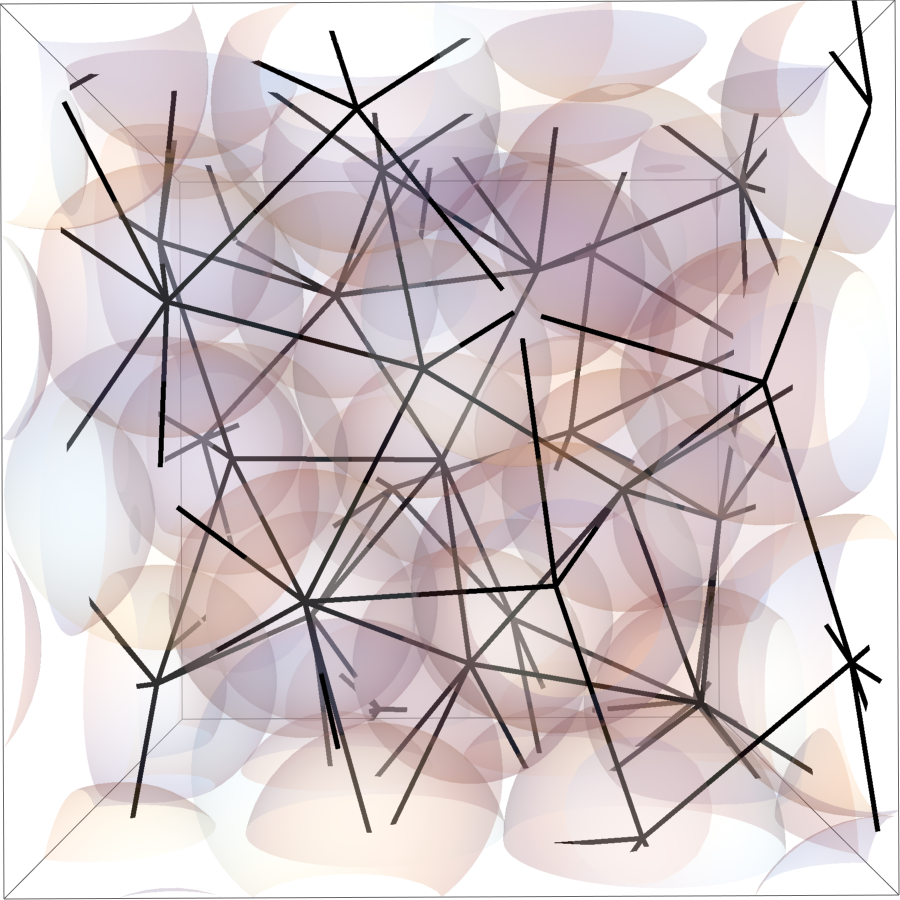}
	\end{subfigure}
	\caption{Left: Jammed configuration of $N=1024$ spheres used to perform the dynamics simulations. Its packing fraction is $\phi_J=0.635$ and each sphere is coloured according to its value of the quantity defined in Eq.~\eqref{def:tot dot CVs}, from light yellow (big values) to dark blue (small, negative values); see also the colour scale in Fig.~\ref{fig:pdf displacement}b. Right: Zoom in a region of the same configuration (without colouring) showing the contact vectors.}
	\label{fig:jammed configuration}
\end{figure*}

For reading ease we summarize here our main results: our starting point is the jammed configuration of spheres shown in Fig.~\ref{fig:jammed configuration} and the resulting network of contact forces. With this information we construct, at the single particle level,
two quantities that are physically well defined --the sum of contact vectors ($\vb{C}$, see Eq.~\eqref{def:tot CV}) and sum of all pairs of dot products between them ($S$, see Eq.~\eqref{def:tot dot CVs})-- and that will be shown to be related to the dynamics.
Before looking for that relation though, we studied the statistical properties of the particles trajectories that occur near the jamming point. We emphasize that this dynamical regime has been little studied so far, specially from a particle-wise perspective (see however \cite{procaccia-agitated-granular,exper_coulais_how_2014}), which justifies our need of first finding a robust characterization of the particles motion. From the results of our numerical simulations, we gather that the statistical distribution of a particle's trajectories can be succinctly described by considering the first moment of its displacement (as a vectorial quantity indicating some anisotropy in the particles motion) and its norm squared (as a measure of its mobility). A sample of these distributions is depicted in Fig.~\ref{fig:pdf displacement} whence we deduce, on the one hand, that some particles have preferential directions of motion, while on the other hand, there is a broad distribution of mobility values, signalling that even close to jamming particles are constrained by their local environment in a heterogeneous way.
We then turned to the question of showing that there is a strong link between these dynamical features and the aforementioned structural variables computed in terms of the contact vectors; see Fig.~\ref{fig:MD-displacements-vs-totCV} (resp. \ref{fig:MC-displacements-vs-totCV}) for the relation with the preferential directions in the Molecular Dynamics (resp. Monte Carlo) simulations and Fig.~\ref{fig:MD-mobility-vs-dotCV} (resp. \ref{fig:MC-mobility-vs-dotCV}) for the connection with mobilities. To further test our approach, we verified that besides finding a significant correlation we can also make some statistical predictions of single particles trajectories in a wide variety of systems by simply ranking the particles according to their value of the structural variables we introduce here(see Fig.~\ref{fig:ranking-forces-and-displacements}). Interestingly, in Fig.~\ref{fig:collisions-vs-t} we provide evidence that the structural information contained in the jammed configuration gets lost rather slowly and \emph{independently} of the dynamical protocols used for the simulations and the different parameters modelling their interactions.
In Sec.~\ref{sec:discussion} we discuss our findings, addressing first the issue of why the most informative structural variables are obtained when the magnitude of the contact forces is ignored. Importantly in Sec.~\ref{sec:comparison-normal-modes} we use the vibrational modes obtained from the Hessian (at jamming) as an alternative scheme to analyse the dynamics of the configuration and find negligible correlations between them. This indicates that, in contrast with our method, the normal modes description fails to capture the statistics of the single-particle trajectories in the dynamical regime we consider here. These results show that the formalism we develop in this work, based solely on exploiting the details of the network of contacts at jamming, provides a more powerful and robust statistical inferential technique than previous works.
Finally Sec.~\ref{sec:conclusions} includes our conclusions and gives some perspectives for future research.

\section{Statics: Physical quantities at the jamming point}\label{sec:statics}

Macroscopically, jammed configurations are usually characterized in terms of their packing fraction, $\phi_J$, whose value depends on some intrinsic parameters of the system, like the number of particles, $N$, as well as particles shapes and sizes. Even for the simple case of spherical particles, the value of $\phi_J$ attained depends on the type of algorithm used to generate the packing as well as the initial condition. Indeed, from a rigorous point of view, the packing fraction does not suffice to determine the properties of jammed states, and in fact systems with the same value of $\phi_J$ may exhibit different degrees of (dis)order in the positions of their particles\cite{torquato_review_jamming}. Conversely, by tuning a macroscopic property, such as the average number of contacts per particle, one can access jammed states within a broad range of values of $\phi_J$\cite{jiao_nonuniversality_2011,hopkins_disordered_2013,torquato_robust_algorithm,donev_comment_2004}.
What is more, it is now clear\cite{jiao_nonuniversality_2011} that even the same algorithm can produce configurations of considerably lower or higher packing fractions, provided some ``regularity'' is manually included in it. And although all these states are perfectly valid instances of jammed packings, \textit{i.e.} systems whose degrees of freedom are blocked, what is most commonly found in nature are packings of random configurations\cite{ohern_jamming_2003,ohern_reply_2004} whose values of $\phi_J$ are, nonetheless, distributed within a narrow range. Computationally, this feature can be closely reproduced and, if an unbiased protocol is utilized, the configurations thus generated show a rather weak dependence on the (possibly random) initial condition, changes of the system's parameters, and even the specific algorithm used\cite{ohern_jamming_2003,parisi_zamponi-review,md-algorithm}. In this scenario, the value of $\phi_J$ is mainly determined by the dimensionality of the system\cite{md-algorithm} and the distribution of the particles sizes. When $d=3$ (the case we will be concerned with in this work), one can safely state that typical monodisperse isostatic jammed packings have $\phi_J \approx 0.64$.

In contrast, from a microscopic point of view and when considering frictionless spherical particles, either soft or hard, a specific realization of a jamming configuration is uniquely identified\cite{jamming_criticality_forces} by the particles positions and radii $\{\vb{r}_i^{(J)},\ a_i \}_{i=1}^N$. These quantities in turn suffice to determine the set of $N_c$ contact forces between them, $\{ \vb{F}_{(ij)} \}_{(ij)=1}^{N_c}$, where $(ij)$ with $i<j$ denotes an ordered pair of  particles in contact. For reasons explained below, from now on we will only consider monodisperse configurations, \textit{i.e.} $a_i:=R_J$ (the radius at jamming) for all $i=1,\dots,N$.
In addition, we will also restrict our analysis to configurations that are isostatic, \textit{i.e.} systems in which the number contact forces matches exactly the number of degrees of freedom, $N_c = N_{dof}$, and consequently are marginally stable\cite{parisi-et-al-jamming-annrev,gardner-perspective_2019,muller_marginal_2015}. These considerations should show that the task of producing a jammed state is clearly not easy because the usual picture of a very rough energy landscape with an exponential number of minima, common to disordered systems, is further complicated in the case of isostatic jamming points due to the marginal stability of each of them. In addition, the need of having direct access to the network of contact forces relevant for our work imposes a requirement that we also need to take into account.
To tackle all these difficulties we implemented a sequential Linear Programming (LP) algorithm\cite{artiaco_baldan_parisi} that produces jammed configurations using as seed spheres at random positions and of small radius. At each step of the algorithm, a non-overlapping constraint is enforced on each pair of particles and their position is updated in such a way that their size can be maximally increased once the spheres have taken the new position. Details about the algorithm can be found in the Appendix \ref{sec:details algorithm}, so here we only report the main properties of the configurations thus obtained. First, since the algorithm stops when no further size increase can be performed because all the degrees of freedom of the system are blocked, a jammed configuration is indeed obtained. More importantly, the dual variables associated with the non-overlapping constraint are equal, up to a proportionality factor, to the magnitudes of the forces exerted between particles in contact.
With these quantities and the positions of the spheres at the jammed state, we were able to construct the full network of contact forces which in turn was used to verify that the final configuration is isostatic, in the sense mentioned above, but also verifying the more stringent condition that the system is posed at mechanical equilibrium, \textit{i.e.} $\displaystyle \sum_{j \in \partial i} \vb{F}_{ij}=0$, where $\vb{F}_{ij}$ is the force acting on particle $i$ due to its contact with $j$, and $\partial i$ is the set of all the neighbours of particle $i$; naturally, $\vb{F}_{ij}=\vb{F}_{(ij)}$ if $i<j$, or $\vb{F}_{ij}=-\vb{F}_{(ij)}$ otherwise. As a further check, we also confirmed that our configurations present all the properties encountered before using other algorithms\cite{wyart_marginal_2012,lerner_low-energy_2013,degiuli_force_2014,hagh_broader_2019} and predicted by the best current available theory based on mean field\cite{jamming_criticality_forces,parisi-et-al-jamming-annrev}; see Appendix \ref{sec:properties jamming}. It is worth noting that all the algorithms produce a small fraction of rattlers (particles with less than $d+1$ in $d$ dimensional systems, so 4 contacts in our case) that should not be considered when counting the degrees of freedom. The reason is that they do not contribute to the global rigidity of the system, \textit{i.e.} the network of contacts would still yield a finite bulk modulus were these particles to be removed. This rigidity along with the fact that our system was generated using periodic boundary conditions imply that the number of degrees of freedom relevant for the contact counting criterion for isostaticity is equal to $N_{dof}= d(N'-1)+1$ \cite{donev_pair_2005,goodrich_finite-size_2012,jamming_criticality_forces}, with $N'\lesssim N$ the number of non-rattlers in the system.

\begin{figure*}[!htb]
	\centering
	\includegraphics[width=1.0\textwidth]{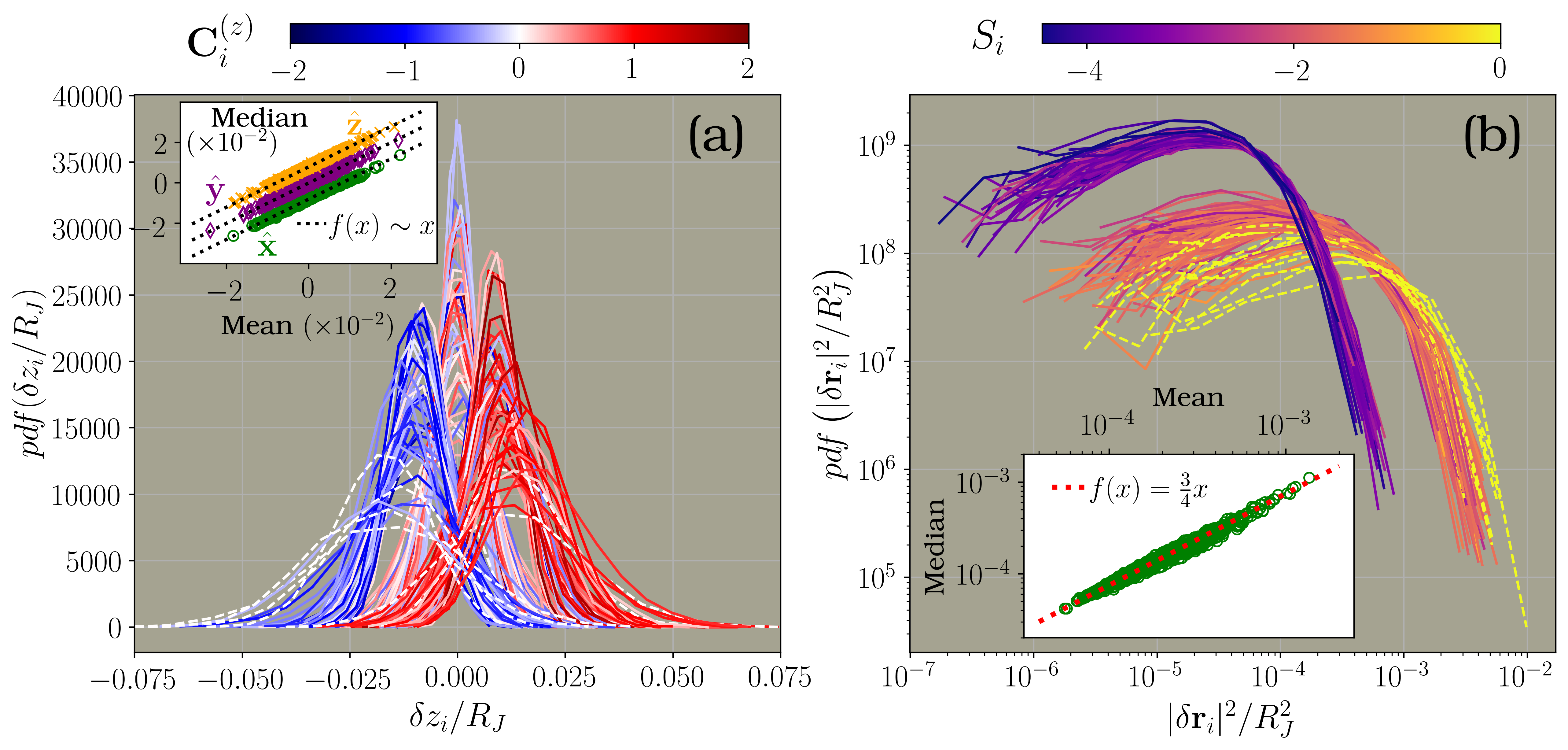}
	\caption{
		Main figures: Probability distributions (pdf) of the variables of Eqs.~\eqref{eq:displacement} at a fixed time $\tau_{MD}=20$ and packing fraction $\phi/\phi_J=0.995$ obtained from $M_{MD}=5000$ independent trajectories sampled from the ICE. Each curve corresponds to the pdf of a single particle, but only $15\%$ of the configuration is shown as described in the main text. (a): pdf of particles' displacement along the $\vu{z}$ direction (similar results hold for the other directions), and (b): pdf of their square displacement. Curves colours correspond to the particles value of $\vb{C}_i^{(z)}$ (in panel a) or $S_i$ (in panel b) as indicated in the scales on top, while dashed lines are used to identify if the curve is associated with a rattler. Insets: Correlation between mean and median of the distributions. In (a) the points of each set have been displaced vertically by a fixed amount for clarity reasons and the identity lines (also displaced by the same amount) are shown in black dotted, while in (b) the red dotted line corresponds to $f(x)=0.75x$.
}
	\label{fig:pdf displacement}
\end{figure*}

With this method we produced monodisperse configurations of $N=1024$ spheres and in Fig.~\ref{fig:jammed configuration} we illustrate\cite{ovito} a typical realization that was afterwards used in the dynamical simulations. In  the left panel, the spheres have been coloured according to their value of $S_i$ defined soon below in Eq.~\eqref{def:tot dot CVs}, while the right panel depicts the resulting contact vectors between neighbouring spheres in a small region of the same configuration. It is important to mention that even though our algorithm can produce polydisperse jammed packings, were we to use spheres with different sizes then the smallest ones would be, presumably, the most mobile ones since they would experience less collisions with their neighbours, rendering them less constrained on average. But this would be a consequence exclusively of their size, and therefore \emph{independent of their surroundings or any structural property of their vicinity}.
Hence, in order to avoid this sort of trivial inference and to really isolate the contribution coming solely from the structure, we restricted our analysis to monodisperse systems. The configuration used in this work and depicted in Fig.~\ref{fig:jammed configuration} has a packing fraction of $\phi_J=\frac{4\pi N R_J^3}{3L^3}=0.635$, with $R_J=0.582$ being the spheres' radius once the jamming point is reached. Only $1.3\%$ of the $N=1024$ particles are rattlers and none of them have any contact forces acting on them. As we will see later, this lack of constraints can be related to the fact that close to jamming most of them are able to move more freely than the rest of the particles.


We now address one of the main goals of our work: how to use the network of contacts formed at jamming to construct well defined physical quantities that can be used as predictors of the dynamics that takes place close to such point. Yet, before answering this question, we want to remark why this approach differs from the previous ones. Most importantly, note that by using a jammed sate as the initial configuration for the dynamics our knowledge is augmented in comparison with the scenario where the initial state is only \emph{close} to a jammed configuration. In this latter case we would be missing data about the true contacts between particles, and thus our description would be limited to the usage of ``coarse-grained'' variables, for instance, the local density or a pair correlation function defined within some small vicinity of each particle, not unlike the scheme of the machine learning methods\cite{schoenholz_combining_2018,schoenholz_identifying_2015,schoenholz_nature_2016,schoenholz_relationship_2017,schoenholz_structural_2016}. In contrast, in the case we consider here the dynamics departs from a configuration in which we can identify precisely, for each particle, which are its true neighbours and, therefore, just include such relevant particles in our description. This distinction is crucial for uncovering preferential directions in the particles' displacements, as we investigate in the next section. Moreover, because in our dynamical simulations we only consider contact potentials between the spheres, a local ``coarse-grained'' description as the one just described would presumably fail to provide a realistic picture of the interactions driving the particles trajectories. In other words, because we will explore the particles' motion that takes place in the vicinity of a jamming point it is to be expected that the correlation between structure and dynamics should be stronger than in other scenarios, and therefore if we have at hand a more accurate information about the structure we will be able to achieve a better and more durable statistical inference of the dynamics.
With this in mind, let us use such network to introduce some quantities that will be used in the rest of this work to describe the statistics of the particles' trajectories. First, we will denote the (unit) contact vector between neighbouring particles $i$ and $j$ as $\vb{n}_{ij} = \frac{ \vb{r}_i^{(J)} - \vb{r}_j^{(J)}}{\abs{\vb{r}_i^{(J)} - \vb{r}_j^{(J)}} }$; they define the edges of the network shown in the right panel of Fig.~\ref{fig:jammed configuration}. Using these vectors, we can easily construct the following two quantities: i) the vectorial sum of all the contact vectors acting on the $i$-th particle,
\begin{equation}\label{def:tot CV}
\vb{C}_i : = \sum_{j \in \partial i} \vb{n}_{ij} \qc
\end{equation}
and, ii) the sum of all pairs of scalar products of the contacts acting on a particle,
\begin{equation}\label{def:tot dot CVs}
S_i := \sum_{j<k \in \partial i} \vb{n}_{ij}\vdot \vb{n}_{ik} = \frac12 \left( \abs{\vb{C}_i}^2 - q_i \right) \qc
\end{equation}
where $q_i$ is the coordination number at jamming of particle $i$. The colours used in the left panel of Fig.~\ref{fig:jammed configuration} indicate the value of $S_i$ for each particle of the system, with a bigger (smaller) value corresponding to a colour on the yellow (dark blue) part of the scale reported in Fig.~\ref{fig:pdf displacement}. The complex distribution of colours displayed by the configuration resembles the ones found using other structural variables or order parameters, which in turn have been linked with the heterogeneous dynamics observed in glassy systems\cite{schoenholz_nature_2016,local_thermal_energy,tong_prx,tong_prl,berthier_static_2012,charbonneau_pts, procaccia-agitated-granular}.

As we will argue in the next sections, it is the contact \emph{vectors} and not the \emph{forces} which actually convey more information about the motion of the configuration near its jamming point. In the Appendix \ref{sec:properties jamming} a more detailed analysis for distinguishing between $\vb{n}_{ij}$ and $\vb{F}_{ij}$ is carried out. It is important to emphasize that both $\vb{C}_i$ and $\vb{S}_i$ are well defined physical variables for each particle of the configuration, and therefore our approach will provide a description of the system's dynamics at the level of individual particles. We thus extend other techniques where the characterization of the particles' displacement was done in terms of clusters or mesoscopic regions within the system.

\section{Dynamics near the Jamming Point}\label{sec:dynamics}

Even though the jamming regime is by itself interesting and complicated enough, many of the most salient physical properties of amorphous solids are linked with the dynamical slowing down that takes place when, say, their density increases. So we now turn to the main part of this work in which we will first characterize the dynamics of our configuration of spheres near its jamming point, which can be attained, for instance, by reducing the system's packing fraction by a small amount and providing the particles with small momenta. Then, in a second stage, we will establish a connection between the sluggish motion of the configuration and its static properties computed at the jamming point. At first sight, it might seem that making such a two-step division of the dynamical analysis is somewhat redundant, since we could proceed instead by directly trying to construct some quantity using the network of contact forces and then correlate it with the heterogeneous dynamics exhibited by glassy systems. This is the most common methodology in studies trying to link the local environment of particles in a configuration with their square displacement, where this last quantity is always taken as a measure of the ``mobility'' of a particle.

As mentioned above, this technique has been used to relate structural features of several model systems with their dynamics.  Nevertheless, given that here we are considering the dynamics in a different and rather unexplored regime, we opted for first trying to answer the question of how particles move near the jamming point. In particular, do they have any preferential direction of motion? And if they do, how much they move along it? To show that these attributes are not necessarily determined by the mobility of a particle, consider the case that its closest neighbours are distributed uniformly around it forming a large cage. It is clear that the particle will be highly mobile, while if the cage was small its motion would be more constrained, and yet, in neither of those two cases, the particle's displacement would exhibit a preferred direction. Conversely, if the arrangement of neighbours is quite irregular, the motion of the particle should be facilitated towards the direction where fewer obstacles are present, however, this favoured path will only be discernible if the particle is also allowed to move enough, because otherwise the preferred direction will be washed out, for example, by thermal noise. These elementary examples illustrate that even a very simple study of the dynamics near jamming should take into account these two properties in order to yield a sufficiently comprehensive picture. Therefore, our approach is based on analysing the particles' displacement both as a vectorial quantity (that is, component-wise) and as a scalar one (using its norm squared, henceforth also termed ``mobility''):
\begin{subequations} \label{eq:displacement}
	\begin{align}
	\delta \vb{r}_i(t) & = \vb{r}_i(t) - \vb{r}_i(0) = (\delta x_i(t), \delta y_i(t), \delta z_i(t) ) \label{seq:displacement}\\
	\abs{\delta \vb{r}_i(t)}^2 & = \delta x_i^2(t) + \delta y_i^2(t) + \delta z_i^2(t) ; \label{seq:square displacement}
	\end{align}
\end{subequations}
where $\vb{r}_i(t)$ is the position of the $i$-th particle at time $t$. Next, we need to probe how the configuration evolves as it explores the phase space close to the minimum determined by the jammed state. Since we want to uncover how the properties of this minimum are connected with the dynamics, it is natural to use as initial condition the jammed configuration, \textit{i.e.} $\vb{r}_i(0) = \vb{r}_i^{(J)}$, and then generate many independent trajectories. This scenario is equivalent to the approach of the iso-configurational ensemble\cite{widmer-cooper_how_2004,widmer-cooper_relationship_2005, widmer-cooper_predicting_2006,widmer-cooper_study_2007,widmer-cooper_irreversible_2008,jack_information-theoretic_2014} (ICE), because we will keep fixed the initial positions in all cases and then let the configuration evolve according to some prescribed dynamical protocol. By sampling from the ICE, we will be able to really isolate the contribution of the network of contact forces to the dynamics because any stochastic contribution to the dynamics is expected to cancel out if the sample of trajectories is large enough. In the next two sections we present the resulting statistics of the particles trajectories in the ICE generated using two different dynamical schemes, namely Molecular Dynamics (MD), where the infinitely hard sphere model is kept, and Monte Carlo (MC) simulations, where we considered soft spheres with different interaction potentials. The reason for using these two types of simulations is that we want to investigate the applicability of our method as we change the model system and its parameters. This is specially important because other studies\cite{hocky_correlation_2014} using the ICE have reported that the level of correlation between local structure and dynamics is very sensitive to the glass former model and its parameters.

\begin{figure*}[h!]
	\centering
	\includegraphics[width=0.99\textwidth]{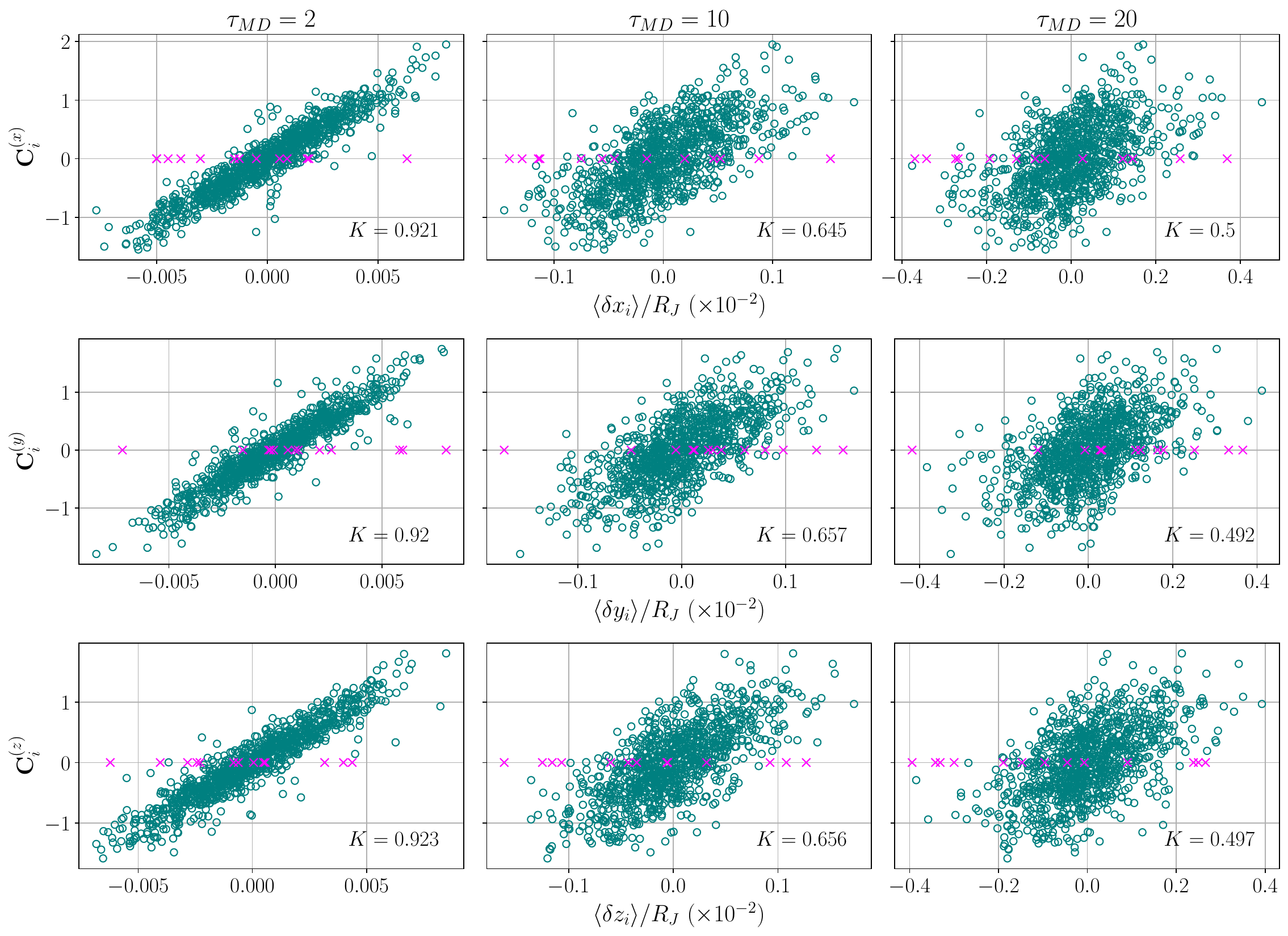}
	\caption{Scatter plots of the particles mean displacement (scaled by the jamming radius $R_J$) and the sum of contact vectors, $\vb{C}_i$, defined in Eq.~\eqref{def:tot CV}. The reported values correspond to ICE averages with $\phi=0.995 \phi_J$. Each row corresponds to one spatial dimension, while each column depicts the values of the mean displacement at different times (measured in collision events, $\tau_{MD}$). Rattlers are identified by pink crosses and the value of the Spearman's rank correlation coefficient is indicated in the lower right part of each panel.
	}
	\label{fig:MD-displacements-vs-totCV}
\end{figure*}

\subsection{Molecular Dynamics simulations}

We begin by showing the results obtained using the MD simulations, for which the radius of all spheres was reduced by the same factor in order to reach a new density $0.9 \leq \phi / \phi_J \leq 0.999$ that was afterwards kept constant. For each value of $\phi$ we then performed $M_{MD}=5000$ independent simulations of event-driven MD using the algorithm described in \cite{md-algorithm}, with initial velocities assigned randomly according to a Maxwell--Boltzmann distribution at inverse temperature $\beta=10$ (in the reduced units of the algorithm) and, as mentioned above, taking $\vb{r}_i^{(J)}$ as initial condition for all the trajectories. More details about these simulations are given in Appendix \ref{sec:supplementary MC}, but here we just mention that we used an event-driven algorithm, so the ballistic regime is absent from all of our data. Additionally, when using this type of algorithms the natural time unit of the dynamics is the number of collisions per particle that have occurred. In order to make a clear distinction between this time unit and the ``physical time'' $t$ in Eqs.~\eqref{eq:displacement} above, we denote as $\tau_{MD}$ the number of events per particle. However, Fig.~\ref{fig:collisions-vs-t} in Appendix \ref{sec:supplementary MD} shows that once we have fixed $\phi$, there is a a well defined relation between $\tau_{MD}$ and $t$.

We will use a statistical characterization of the particles' motion, which we can access after generating the $M_{MD}$ trajectories in the ICE for a fixed packing fraction. The resulting probability distribution function (pdf) of the $\vu{z}$ component of the particles displacement at $\tau_{MD}=20$ and using $\phi=0.995\phi_J$ is shown in Fig.~\ref{fig:pdf displacement}a (the distributions of the other components are very similar). Keeping in mind that each curve is the pdf of a single particle, it is clear that all the particles have a well defined mean displacement, many of which are \emph{different from zero}. This is the first of our main results, because it indicates that some spheres indeed have a preferred direction of motion and that it can be identified despite the statistical fluctuations coming from the thermal noise and sample-to-sample variations. More importantly, we can infer this particular direction using the sum of contact vectors defined in Eq.~\eqref{def:tot CV}, as can be seen from the evident division of colours to the left and to the right and the scale at the top of the figure. Only $15\%$ of the particles were used to generate the main figure, and for clarity reasons they where chosen as the ones having the largest absolute value of $\avg{\delta z_i}$ ($10\%$) and the remaining $5\%$ as the ones with the smallest absolute value, but no information of their distributions width was used whatsoever. We emphasize here that all the mean values and other statistics that will be reported refer to ICE averages. Since all the distributions shown are unimodal and rather narrow, we can expect the value of $\avg{\delta z_i}$ to be descriptive enough. To test this idea, we compared the mean value and the median of each particle in the configuration and obtained the results included in the inset of the same figure, where data points of different colour correspond to the different spatial components as indicated. The fact that all of them lie very close to the identity line (black dotted curves) confirms that $\{ \avg{\delta \vb{r}_i} \}_{i=1}^N$ can be used to discern the existence of preferential directions in the possible trajectories of the configurations. We note in passing that the presence of anisotropy in the particles' motion has only been recently studied in few works, \textit{e.g.} \cite{rottler_predicting_2014,patinet_connecting_2016,xu_predicting_2018,barbot_local_2018,schwartzman_anisotropic_2019}.


\begin{figure*}[htb!]
	\centering
	\includegraphics[width=0.99\textwidth]{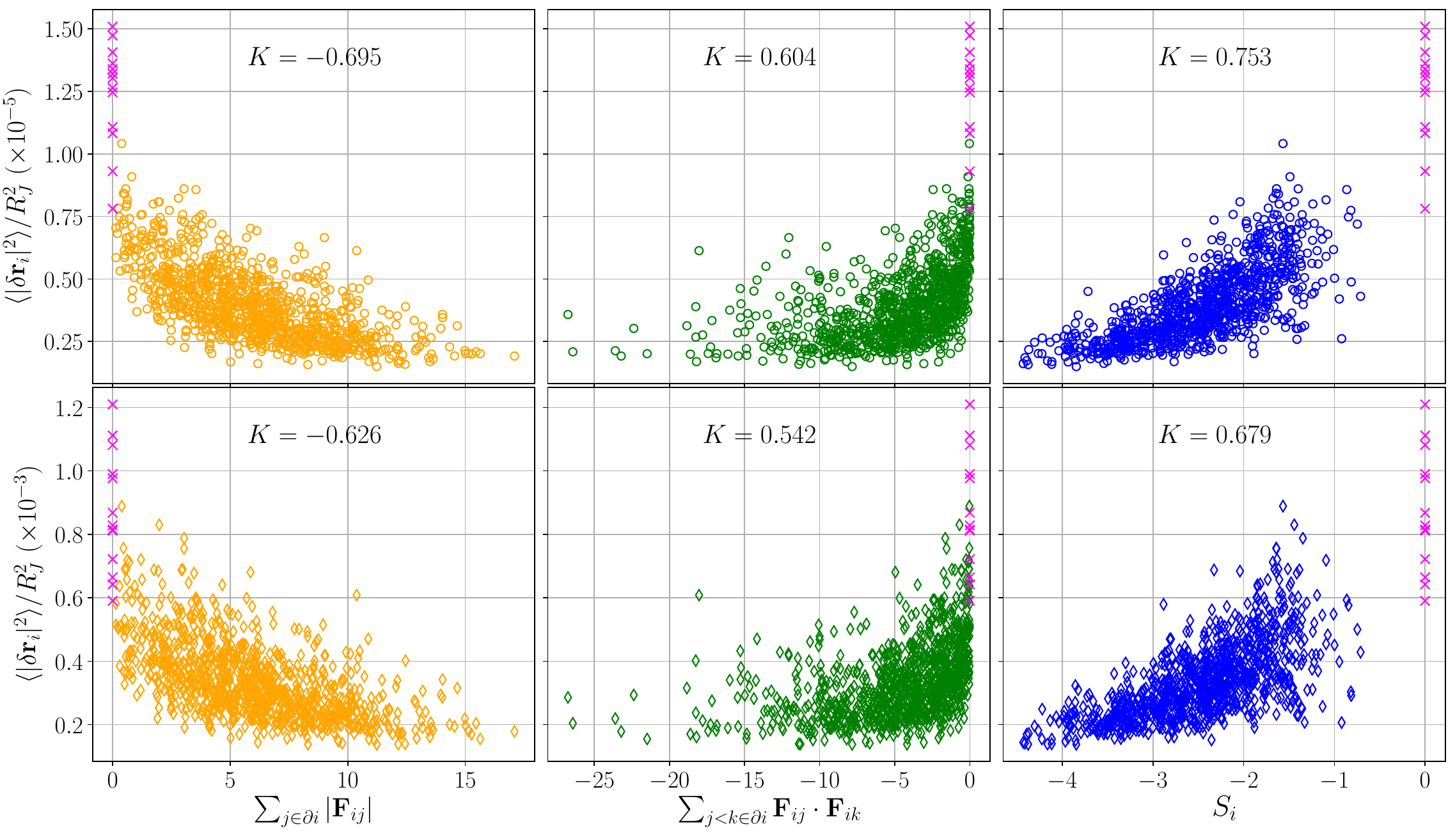}
	\caption{
	Scatter plots of the particles' average mobility and three possible scalar quantities that can be constructed using the network of contact forces at the jamming point: the sum of forces magnitudes (first column), the sum of dot product between pairs of contact \emph{forces} (second column), or using the contact \emph{vectors} (third column) as defined in Eq.~\eqref{def:tot dot CVs}. In the upper panels we report the results from simulations with density $\phi/\phi_J=0.999$, while the lower ones show the analogous results using $\phi/\phi_J=0.99$. As in the previous figure, rattlers are also identified with pink crosses and the Spearman's rank correlation coefficient is also included.
			}
	\label{fig:MD-mobility-vs-dotCV}
\end{figure*}

Analogous results for the particles' square displacement are shown in Fig.~\ref{fig:pdf displacement}b, also using $15\%$ of the particles, chosen with the same criterion as before but this time taking into account the mean value of their mobility; correspondingly, the colour scale indicates the value of $S_i$ defined in Eq.~\eqref{def:tot dot CVs}.
Importantly, in this case it should be noticed that the shape of the distributions is notoriously different with respect to the ones of $\delta \vb{r}_i$. This is exemplified by the fact that the mean and median of the mobility do not coincide in this case (see the inset) because the mean is a statistic more sensible to extreme events, identified with trajectories where the mobility can be about 10 times larger than the most typical ones. Nevertheless, there is a simple linear relation between both statistics, as illustrated in the inset where the red dotted line indicates the curve $f(x)= \frac{3}{4}x$.
In summary, the distributions of $\delta \vb{r}_i$ and $|\delta \vb{r}_i|^2$ have clearly a different shape, with unimodal, well peaked curves for the former, and broader, asymmetric probability densities in the latter case. Such difference evinces that these two variables have different behaviours and should therefore be studied separately as we will do here. We also remark that, by the same token, it is not obvious that the ICE mean of a particle's mobility, \textit{i.e.} $\avg{|\delta \vb{r}_i|^2}$, would be a good descriptor of its full distribution.  Therefore, to justify the usage of such average to characterize the statistics of $|\delta \vb{r}_i|^2$, we provide the following reasons:
i) the support of the distributions shown in the figure differs, roughly, by an order of magnitude, which is enough to tell apart the least mobile particle from the most mobile ones; ii) the inset shows that there is still a linear relation between the median and the mean, implying that a higher value of the latter is accompanied by a proportional translation of the full distribution also to higher values; iii) other studies using the ICE have found that several dynamical features worth studying in glassy systems are indeed captured by $\avg{|\delta \vb{r}_i|^2}$.
In any case, a word of caution is in place to avoid possible confusions regarding how to interpret the mobility distributions. First of all, one should not expect the first moment of $|\delta \vb{r}_i|^2$ to reflect its typical values, as has already been reported in \cite{berthier_structure_2007}. Additionally, the grouping of the curves depicted in the main panel of Fig.~\ref{fig:pdf displacement}b in two different sets should not be understood as if there were two limiting distributions of the mobilities, but instead as a consequence of the fact that we selected only $15\%$ of the particles trajectories; that is, the remaining $85\%$ of the curves fill the gap between the apparently two different sets.

We now turn to the question of finding how much information the jamming point contains about the dynamics close to it. This means that the connection between similar local environment and similar dynamics, illustrated in Fig.~\ref{fig:pdf displacement} for few particles by the clustering of distributions of similar colours, must now be extended to the full configuration.
A convenient way to achieve this is to use just few statistics of the distributions, instead of the complete pdf. As argued above, the first moment comes across as a quantity descriptive enough and the scatter plots of Fig.~\ref{fig:MD-displacements-vs-totCV} confirm that there is indeed a correlation between $\avg{\delta \vb{r}_i}$ and the sum of unit contact vectors, $\vb{C}_i$ defined in Eq.~\eqref{def:tot CV}. Each of the three rows in the figure corresponds to a single direction, while the different columns refer to the different times of the dynamics at which the mean displacement was calculated. Once again, we note that the distribution of values of the mean displacement is broad enough to conclude that many of the particles have a preferential motion direction, \textit{i.e} $\avg{\delta \vb{r}_i} \neq 0$. It is worth emphasizing that since we used a large set of trajectories this latter fact cannot be attributed to statistical noise. Another relevant feature that validates our approach is the fact that the rattlers (shown with pink crosses in Fig.~\ref{fig:MD-displacements-vs-totCV} and whose pdf are drawn with dashed lines in Fig.~\ref{fig:pdf displacement}) generically exhibit a larger value of mean displacement, which is consistent with the claim that their trajectories are less constrained due to a smaller amount of nearby particles. Expectedly, it is clear that initially the dynamics of each particle is highly correlated with its value of $\vb{C}_i$, but that as more collisions occur this information eventually gets lost.

These results corroborate that we can use our knowledge about the contact vectors to identify preferential directions in the motion of individual particles. To measure the quality of such an inference we used the Spearman's rank correlation coefficient, $K$, whose value is reported in the lower-right part of each panel. The advantages of using $K$ for studying the correlation between the local structure and the dynamics of glassy systems are twofold, since for one part this coefficient has proven to be very sensitive to the effect of varying the different parameters of a glass former model\cite{hocky_correlation_2014}, while on the other it reflects naturally our inferential approach based on the hypothesis that the ranking of particles by a suitably chosen static variable (in this case $\vb{C}_i$) should have a direct connection with the ranking obtained via a dynamical variable (here $\avg{\delta \vb{r}_i}$).


The influence of a particle's set of contact vectors in establishing its preferred direction of motion can be understood because $\vb{C}_i$ provides a good description of where the particle's neighbours are located and hence which are the directions that will be favoured by possible collisions. More precisely, a large value of, say, $\vb{C}_i^{(x)}$ indicates that the particle's first neighbours are arranged in such a way that their net effect is to predominantly push it in that direction, while a mostly uniform arrangement will lead to a very small value of $\vb{C}_i$ and thus the particle would mainly remain in a small vicinity. On the other hand, we do not expect $\vb{C}_i$ to appropriately describe the fluctuations of a particle's motion around its mean displacement mainly because the information about the directionality is lost when computing the second moment of its displacement, and this is indeed what we found when a similar analysis is done for $Var[\delta \vb{r}_i] = (Var[\delta x_i], Var[\delta y_i], Var[\delta z_i])$ as shown by the Fig.~\ref{fig:MD-variance-vs-totCV} in Appendix~\ref{sec:supplementary MD}.

To continue, let us recall that a particle's squared displacement, which does not necessarily incorporate any inherent directionality of the motion, is an important feature to take into account for describing its dynamics in the high density regime mainly because if provides an estimate of the ``cage'' size in which the particle is moving. As anticipated in Fig.~\ref{fig:pdf displacement}, there is a close correlation between $\avg{|\delta \vb{r}_i|^2}$ and $S_i$, however there is no reason \emph{a priori} why this variable should be used instead of other scalar observables that can be obtained from the network of contacts, in particular, others that take into account the forces magnitudes.
To verify that this is not the case and that the direction of contacts provides a better prediction of the particles' mobility, we considered here two other scalar quantities that, on the one hand, are physically well defined and that can be easily computed from the network of contacts at jamming, while, on the other, should be intuitively related to the statistics of $\abs{\delta \vb{r}_i}^2$: i) the sum of all the contact forces magnitudes, $\displaystyle \sum_{j \in \partial i} \abs{\vb{F}_{ij}}$; ii) the sum of all the pair of dot products between contact \emph{forces}, $ \displaystyle \sum_{j<k \in \partial i} \vb{F}_{ij}\cdot \vb{F}_{ik}$. In Fig.~\ref{fig:MD-mobility-vs-dotCV} we compare the correlation between these two observables (first two columns) as well as $S_i$ (right-most column) and $\avg{|\delta \vb{r}_i|^2}$, calculated at $\tau_{MD}=10$. Upper (resp. lower) panels show the results from the MD trajectories with a density of $\phi/\phi_J=0.999$ (resp. $\phi/\phi_J=0.99$), as before data associated with rattlers are also indicated with pink crosses and the values of $K$ are also included for comparison. As expected, in all cases the correlations attained in the system of higher density are larger because the local structure resembles more the one of the jammed state and hence the dynamics is more influenced by the contacts with the closest neighbours at such state. Yet, at first sight it seem paradoxical that including more information, namely, the magnitudes of the forces actually diminishes the predictive power. However, before explaining why this happens, we turn to present the results obtained with the other dynamical protocol, where very similar results were found, and hence a common explanation apply to both types of simulations as will be discussed in Sec.~\ref{sec:discussion}.
Finally, we point out that we tested this same methodology on another independent jammed configuration as described in the Appendix \ref{sec:second-configuration} obtaining essentially the same results, see Figs.~\ref{fig:MD-displacements-vs-totCV-2nd-config} and \ref{fig:MD-mobility-vs-dotCV-2nd-config}.

\subsection{Monte Carlo simulations}

In contrast with the MD simulations, for the MC ones we fixed the spheres' radius at $R_J$ and soften the interaction between them by introducing a contact potential between particles pairs of the form
\begin{equation}\label{def:potential soft sphere}
U(\vb{r}_i, \vb{r}_j) = \abs{2R_J - \abs{\vb{r}_i - \vb{r}_j}}^\nu \Theta\qty(2R_J - \abs{\vb{r}_i - \vb{r}_j}) \, ;
\end{equation}
where $\Theta$ is the Heaviside step function and $\nu$ plays the role a ``stiffness'' parameter that we varied to probe the effect of different types of interactions between the spheres. Analogously to the MD simulations, we used the jammed configuration as initial condition to generate a trajectory using the Metropolis-Hastings algorithm at a fixed temperature $T$. Hence, for this type of dynamics, the natural time unit is the number of MC steps performed, which we will denote as $\tau_{MC}$ to distinguish it from the time scale of the MD simulations and the physical time. We tried different interaction potentials by setting $\nu=\{3/2,2,5/2\}$ (henceforth referred as sub-harmonic, harmonic, and Hertzian interaction, respectively) as well as several values of $T$, performing $M_{MC}=1000$ MC simulations for each value of these parameters. The different values of $T$ were selected in such a way that the samplings were done with a similar acceptance rate; this and other details of the simulations are provided in the Appendix \ref{sec:supplementary MC}.

\begin{figure*}[htb!]
	\centering
		\includegraphics[width=0.95\textwidth]{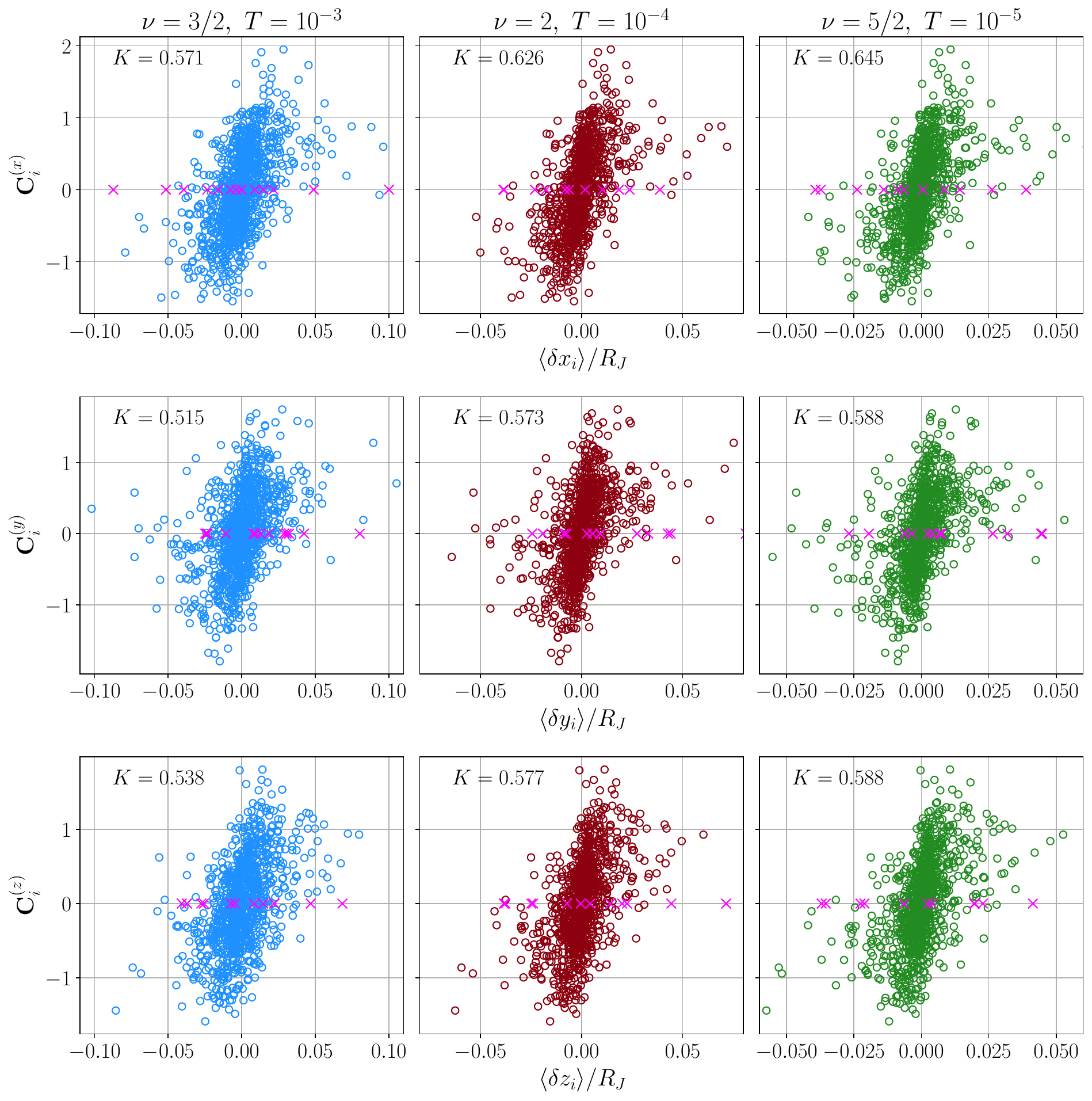}
	\caption{
		Scatter plots of $\avg{\delta \vb{r}_i}$ and $\vb{C}_i$ (each row corresponds to a component), and for the three types of interaction considered: sub-harmonic (first column), harmonic (second), and Hertzian (third); different colours are included just for clarity sake. The temperature used to generate the MC trajectories for each potential is indicated at the top, and the values of $\avg{\delta \vb{r}_i}$ reported correspond to $\tau_{MC}=250$ MC steps. For each value of $\nu$, the corresponding values of temperature were chosen so that the average energy of the system divided by $T$ remained within a small range; see also Appendix \ref{sec:supplementary MC}. As in the analogous plots of MD simulations, the quantities associated with rattlers are identified by pink crosses and the value of $K$ is also included.
	}
\label{fig:MC-displacements-vs-totCV}
\end{figure*}

\begin{figure*}[htb!]
	\centering
	\includegraphics[width=0.95\textwidth]{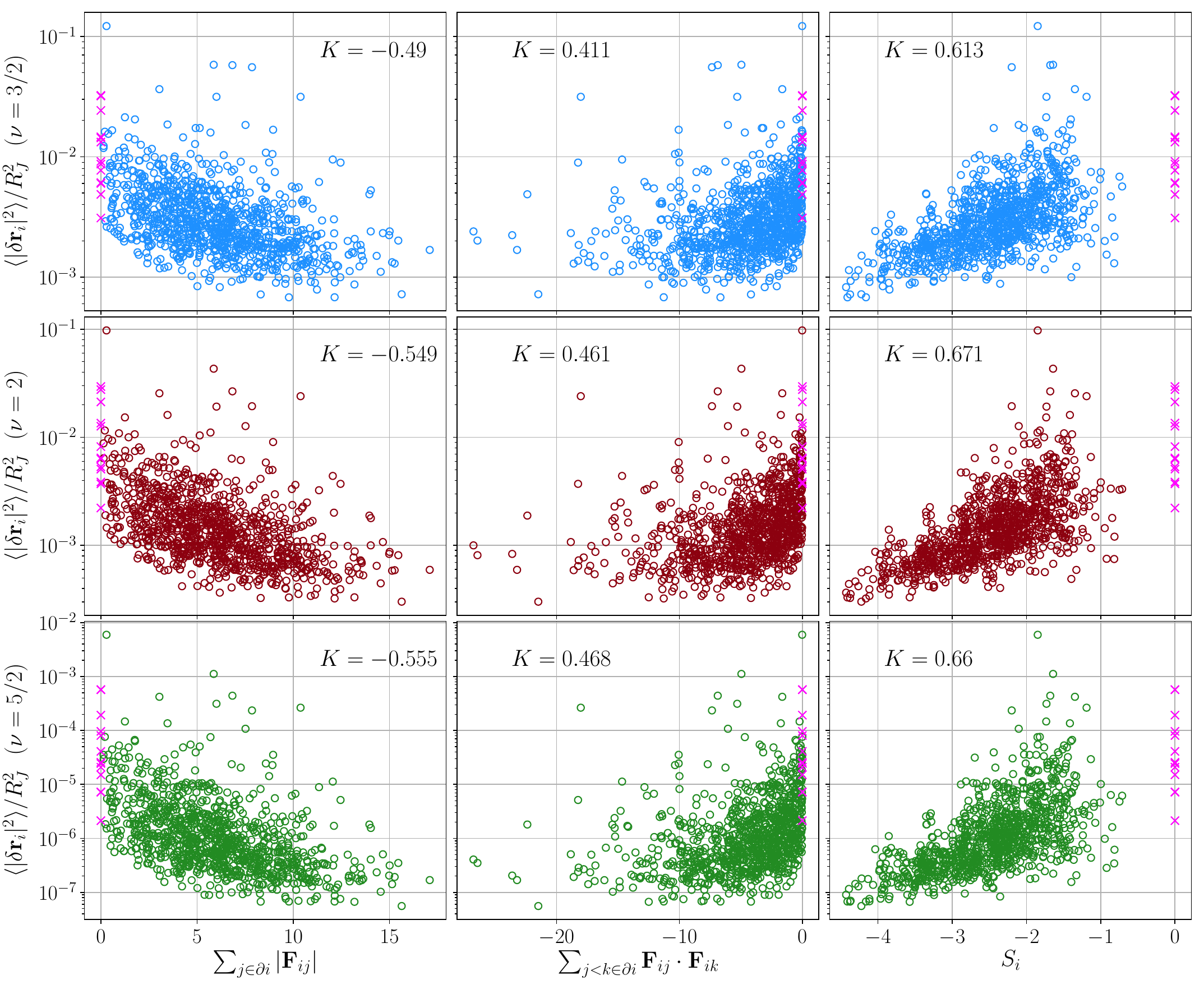}
	\caption{Scatter plots of each particle's average mobility at $\tau_{MC}=250$ with the same interaction parameters and temperatures of Fig.~\ref{fig:MC-displacements-vs-totCV}. Each row consists on the same values of $\avg{|\delta \vb{r}_i|^2}$ as function of different scalar quantities that can be constructed using the network of contact forces at the jamming point, like the sum of forces magnitudes (first column), the sum of dot product between pairs of contact forces (second column), or using the contact \emph{vectors} (third column) as defined in Eq.~\eqref{def:tot dot CVs}. As in the previous figures, points associated with rattlers are identified with pink crosses and the Spearman's rank correlation coefficient is also included.
		}
\label{fig:MC-mobility-vs-dotCV}
\end{figure*}

We now show that with this rather different dynamical protocol we are able to recover very similar results than those of the MD simulations. In Fig.~\ref{fig:MC-displacements-vs-totCV} we compare the correlations between $\avg{\delta \vb{r}_i}$ and $\vb{C}_i$ for the three types of interactions (one in each column) and with the temperatures indicated on top. Each row corresponds to a component of $\avg{\delta \vb{r}_i}$, computed at $\tau_{MC}=250$ in all cases, and for clarity reasons, a different colour is used for each value of $\nu$, while in all panels rattlers are also identified by pink crosses as in the previous figures. The values of $K$ are comparable for all the interaction potentials tested and also with the ones obtained in the MD simulations, thus signalling that our approach is robust enough to be applied to both types of dynamics. It is also worth mentioning that there is a systematic increase of the correlation with $\nu$, since the smallest value of $K$ was obtained with the sub-harmonic interactions, while the Hertzian ones yielded the largest values, independently on the components. However, this last result is related with the fact that different values of $\nu$ produce different decorrelation rates if $\tau_{MC}$ is used as time unit instead of considering a measure of how much the configuration has changed.
In Sec.~\ref{sec:correlation vs time} we show that the temporal evolution of $K$ follows an universal behaviour if such a measure is employed.
Furthermore, also in this case we found that $\vb{C}_i$ is uncorrelated with the variance of $\delta \vb{r}_i$ (see Fig.~\ref{fig:MC-variance-vs-totCV} in Appendix \ref{sec:supplementary MC}) , in agreement with the results of MD. More importantly, when considering the mean mobility of particles and its correlation with the scalar observables constructed from the network of contacts, we also found that $S_i$ is the most informative variable in comparison with the ones that include the magnitudes of the contact forces. This finding holds independently of the type of interaction as shown in Fig.~\ref{fig:MC-mobility-vs-dotCV}, where each column compares the correlation between $\avg{|\delta \vb{r}_i|^2}$ and the three scalar quantities considered before, while different rows (from top to bottom) correspond to the different interaction types (from $\nu=3/2$ to $\nu=5/2$). Once again, we obtained values of $K$ comparable to the ones of the MD simulations, although this time the highest value of $K$ was obtained with the harmonic interaction between particles. But this is due to the fact that there is a different decorrelation rate between different statistics, as we show in Sec.~\ref{sec:correlation vs time}.

To close this section we want to emphasize, that the quantities used here, namely $\vb{C}_i$ and $S_i$, are computed from the properties of the configuration at jamming and therefore are strictly static, nevertheless they can be utilized as descriptors of the statistics of dynamical features of the particles trajectories.

\begin{figure*}[htb!]
	\centering
	\includegraphics[width=0.99\textwidth]{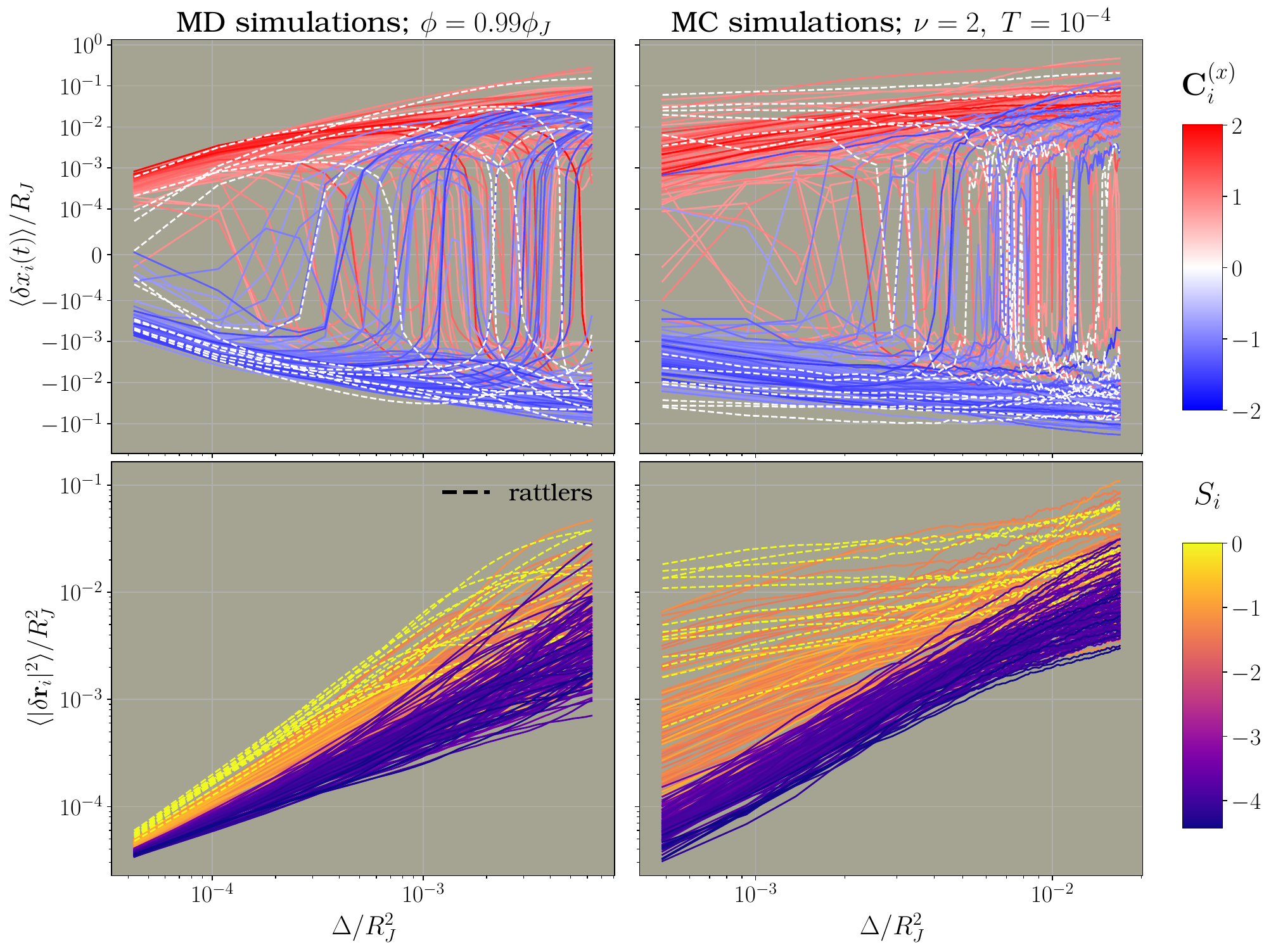}
	\caption{
	Evolution of the $\vu{x}$ component of the mean displacement (upper panels) and the average squared displacement (lower panels) of $20\%$ of the particles for the two type of dynamics utilized here (MD on left panels and MC on the right ones), with the parameters specified on top. The trajectories on the upper (resp. lower) panes were selected by choosing the top and bottom $10\%$ of the particles ranked according to their value of $\vb{C}_i$ (resp. $S_i$); rattlers were also included in the first case. Trajectories are coloured using the scaled shown in the right and the sharp division of colours demonstrates that we can indeed identify both: which particles are the most mobile ones and if they have a preferred direction of motion. We used the mean squared displacement, $\Delta$, to measure the dynamical evolution for the reasons explained in the main text.
	}
	\label{fig:ranking-forces-and-displacements}
\end{figure*}

\section{Prediction and decorrelation from the jammed state} \label{sec:correlation vs time}

 \begin{figure*}[htb!]
 	\centering
 	\includegraphics[width=0.95\textwidth]{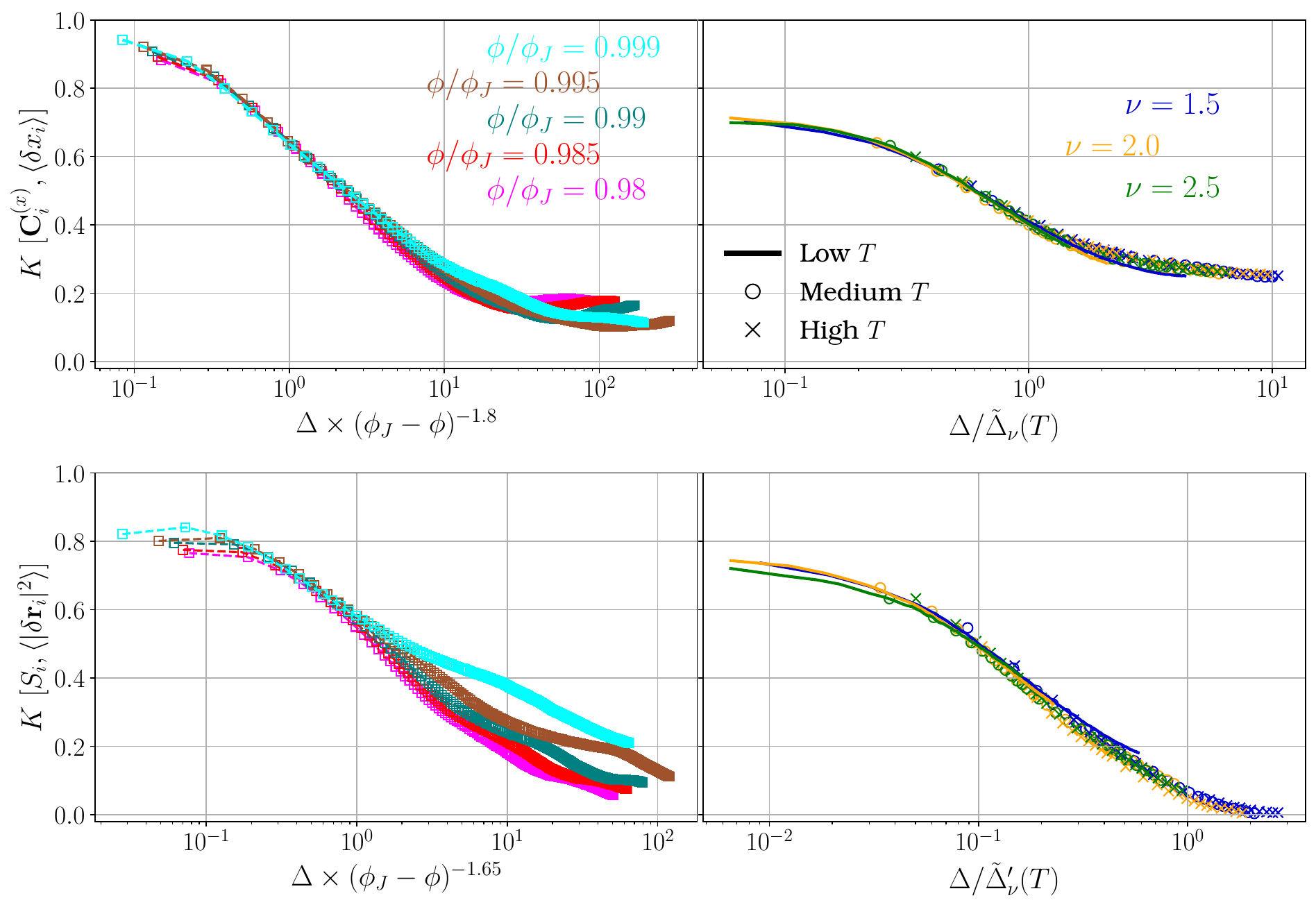}
 	\caption{
 	Evolution of the Spearman's rank correlation coefficient as the configuration moves away from its initial state, measured by the mean squared displacement $\Delta$. We found considerably high values of $K$ for both types of features studied here, namely preferential direction of motion (upper panels) and mobility (lower ones). Our characterization is robust enough to be applied to both dynamical protocols: MD simulations, shown in the left panels (different colours correspond to different packing fraction), and MC simulations (each colour corresponds to a different interaction potential and each line style to a different temperature). Furthermore, we can make the different curves collapse on top of each other when $\Delta$ is scaled as a function of how far from the jamming point the system was posed when generating the trajectories, namely the difference $\phi_J-\phi$ for the MD simulations and $T$ for the MC ones.
 	}
 	\label{fig:correlations-vs-t}
 \end{figure*}

As mentioned in the introduction, there have been many proposals to address the problem of the connection between local structure of disordered solids and their dynamics. However, one of the most pressing issues is not just to find such correlations between statical and dynamical properties, but instead using the former to \emph{predict} the latter. To the best of our knowledge, just a handful of works have dealt in detail with
this problem using well established physical variables\cite{widmer-cooper_predicting_2006,candelier_spatiotemporal_2010}, although more recently in \cite{schoenholz_nature_2016,schoenholz_combining_2018,schoenholz_identifying_2015,schoenholz_relationship_2017,schoenholz_structural_2016,schoenholz_heterogeneous_2019} researchers addressed similar questions employing machine learning methods, obtaining outstanding prediction power despite relying on an artificial representation of the particles' environment at the \emph{mesoscopic} scale, which is an important difference with the approach we have adopted here. At any rate, as a further benchmark of our method we turn next to this point.
Note that this is the inverse scenario of the one used so far, particularly concerning the results of Fig.~\ref{fig:pdf displacement}, because in that case we began distinguishing the particles with some prior information about the dynamics (\textit{i.e.} the ranking based on the first moments of $\delta \vb{r}_i$ and $\abs{\delta \vb{r}_i}^2$) and showed that this distinction carried over to the structural features we considered. In contrast, here we focus on the reciprocal problem where our starting point will be the static information (namely, the contact vectors) which will then be used to infer a property about the dynamics of the configuration. We will show that within our approach, prediction is indeed possible insomuch as we can use $\vb{C}_i$ to reveal preferential direction in the particles' motion, while $S_i$ can be utilized to identify the most mobile ones, \emph{without any prior information about the dynamics}. As expected, the validity of such predictions is highest at the initial part of the trajectories and later it decays as time goes  by and the system evolves. In any case, to be more precise about how the initial correlation is lost we need to find a way to compare consistently the evolution of the configuration in both types of dynamical protocols and for different values of the parameters. A natural candidate is the (averaged) mean square displacement (MSD), $\Delta = \avg{ \frac1N \sum_{i=1}^N \abs{\delta \vb{r}_i(t)}^2 }$, which can be used as a measure of how much the configuration has changed from its initial state determined by $\{ \vb{r}_i^{(J)}\}_{i=1}^N$.
Because we focused on the short-time dynamics, the values attained by the MSD during our simulations are well below the one of its characteristic long time plateau and identified with the onset of the Debye behaviour in the density of states\cite{dynamic-criticality-jamming}. The temporal evolution of $\Delta$ in the MD simulations shown in Fig.~\ref{fig:MD-msd-vs-t} in Appendix \ref{sec:supplementary MD} supports this claim, and it also verifies that all our simulations were done in the same dynamical regime, where the MSD have a similar behaviour for the different packing fractions we used.
 Thus, we will henceforth report the evolution of the quantities we analyse as a function of $\Delta$ instead of as a function of time, $\tau_{MD}$, or $\tau_{MC}$.

 %

We can pose the problem of predicting the preferred directions in configuration space as investigating whether particles for which the observable $\vb{C}_i$ has a large absolute value in one of its components also exhibit trajectories that move predominantly along that same component. We can answer in the affirmative this question following a straightforward method: we rank the particles according to their value of $\vb{C}_i^{(x)}$, then select the top and bottom $10\%$, and plot the evolution of their trajectories. This results in the plots in the upper panels of Fig.~\ref{fig:ranking-forces-and-displacements} for the $\vu{x}$ component of the trajectories generated via the MD simulations with $\phi/\phi_J=0.99$ (left panel) or the MC algorithm with harmonic interaction and $T=10^{-4}$ (right panel). In both cases, the lines colour indicates the value of the sum of contact vectors according to the scale on the right. These figures show that most particles which share a close value of $\vb{C}_i$, and thus whose respective curves are shown in a similar colour, move in the same direction. Trajectories of rattlers are also included (as dashed lines) and even though we can not predict any preferred direction in their motion, we can nevertheless identify them as highly mobile particles  and thus can be used as an estimate of bounds delimiting the maximum displacement of individual particles.
Analogously, if we now proceed to rank the particles according to their value of $S_i$ we have enough information to identify the most mobile particles, as demonstrated in the lower panels of Fig.~\ref{fig:ranking-forces-and-displacements}. In this case, the effect of the different protocols for the simulations is more evident for the initial part, since the MD simulations show a much narrower range of values of the squared displacement, while the MC ones yielded values of $\avg{\abs{\delta \vb{r}_i}^2}$ that differ by more two orders of magnitude. However, the grouping of trajectories with similar colour is still preserved, thus confirming that $S_i$ can be used to distinguish between mostly mobile and mostly blocked particles. Note that in this case, rattlers also provide a good estimate for an upper bound on the value of $\avg{\abs{\delta \vb{r}_i}^2}$ and that only for these particles $S_i$ is identically zero, while for the others $S_i \simeq -1$ at most. As anticipated, the clear division of colours, reflecting an almost perfect correspondence between contact vectors and the statistical properties of the dynamics, is subsequently lost as the configuration moves away from its initial state, or, in other words, as $\Delta$ gets larger.

Results so far show that significant correlations exist between the static features of the configuration and its dynamics near the jamming point. Nonetheless, it is clear that as time evolves such initial correlation should eventually disappear, and therefore a crucial question is how fast and in which way this initial information is lost. As expected, different systems and different parameters yield diverse curves for the temporal evolution of $K[\vb{C}_i, \avg{\delta \vb{r}_i}]$ and $K\qty[S_i, \avg{\abs{\delta \vb{r}_i}^2}]$, where for brevity, we introduced the notation $K[X,Y]$ to denote the Spearman's rank correlation coefficient between variables $X$ and $Y$. Surprisingly enough, when considering the correlation as a function of $\Delta$, and rescaling this latter quantity in terms of the natural variables of the system, we found that all the curves can be made to collapse on top of each other for a significant fraction of the dynamics, signalling the existence of a common process by which the configuration decorrelates from its initial state. Results for $K[\vb{C}_i^{(x)}, \avg{\delta x_i}]$ are shown in the upper panels of Fig.~\ref{fig:correlations-vs-t}, with the results of MD simulations in the left panel and using several values of $\phi$ (each one in a different line colour), while the right one displays the MC results using the three types of interactions (different colours), each one at three different temperatures (different line style). The natural parameter of the MD simulations is the packing fraction of the system, and indeed we found the appropriate rescaling of $\Delta$ to be $(\phi_J-\phi)^{1.8}$. On the other hand, in the MC case each value of $\nu$ sets an intrinsic softness to the spheres, while the temperature determines how much they are allowed to overlap on average. Hence, once the interaction parameter $\nu$ is fixed, the rescaling of the MSD should only depend on $T$, and indeed, we can define a characteristic square displacement $\widetilde{\Delta}_\nu(T) = c_\nu T^{\gamma_\nu}$, where $c_\nu$ and $\gamma_\nu$ are constants that only depend on the interaction type; we report the latter in the first row of Table \ref{tab:values-gammas}. Note that in both cases, the rescaling is essentially a measure of how far away from its athermal jamming point we have posed the system, either by reducing its packing fraction (in the hard sphere scenario) or by providing some thermal energy (in the case of a contact potential).

The lower panels of Fig.~\ref{fig:correlations-vs-t} confirm that very similar findings hold for $K[S_i, \avg{\abs{\delta \vb{r}_i}^2}]$, the only difference being that in this case the best scaling of $\Delta$ in the MD simulations, at least for big enough $K$, turns out to be $(\phi_J-\phi)^{1.65}$, while in the soft sphere scenario of the MC simulations the same type of rescaling applies, albeit with different constants, \textit{i.e.} $\widetilde{\Delta}'_\nu(T) = c'_\nu T^{\gamma'_\nu}$; the values of these exponents are reported in the last row of Table \ref{tab:values-gammas}. We can understand  that the correlations found from $\vb{C}$ and $\vb{S}$ yield different scalings because each of these two static observables provides information about different features of the dynamics, namely the directions in which the displacement of a given particle is facilitated and how much it is expected to move, respectively. It should be noted that the decorrelation occurs rather slowly as revealed by the fact that during the interval in which the dynamical evolution of the configuration makes the value of $K$ to decrease by roughly an order of magnitude, the (rescaled) MSD has increased by at least two orders of magnitude. Collapse of the curves on top of each other suggests that the mechanism driving the loss of correlation with the original basin is independent of the values of the parameters involved in the simulations or even the type of dynamical protocol followed.
We close this section mentioning that we also verified that the same results hold for the other configuration we tested, as shown in Fig.~\ref{fig:correlations-vs-t-2nd-config} in the Appendix \ref{sec:second-configuration}.

\begin{table}[htb!]
\centering
\begin{tabular}{|c|c|c|c|}
\hline
  & $\nu=3/2$ & $\nu=2$ & $\nu=5/2$\\[2mm]
 \hline
 & & & \\
 $\gamma_\nu$ & $0.84$ & $0.64$ & $0.51$ \\[3mm]
 $\gamma_\nu'$ & 0.45 & 0.40 & 0.35\\
 \hline
\end{tabular}
\caption{Values of the exponents used for the rescaling of the MSD with the temperature in the MC simulations, producing the curves shown in Fig.~\ref{fig:correlations-vs-t}. Values in the first row correspond to the results of $K[\vb{C}_i, \avg{\delta \vb{r}_i}]$ while in the second row we report the associated values for $K[S_i, \avg{\abs{\delta \vb{r}_i}^2}]$. }
\label{tab:values-gammas}
\end{table}

\section{Discussion}\label{sec:discussion}

We have shown that the structural variables analysed in this work are strongly correlated with the dynamics of a configuration near its jamming point (Sec.~\ref{sec:dynamics}) and that the decorrelation between the dynamics and the network of contacts occurs in a low, logarithmic fashion (Fig.~\ref{fig:correlations-vs-t}), on the one hand, while it is very robust against changing the interaction potential of the system on the other, (Sec.~\ref{sec:correlation vs time}). Yet, the fact that using the forces magnitudes is detrimental to the inference remains to be explained, which is the issue we will tackle now. After that, we continue this section by putting our results in perspective with previous works, specially Refs.~\cite{exper_coulais_how_2014} and \cite{dynamic-criticality-jamming}. Finally, inspired by this latter reference, we present a detailed exploration of an analogous description of the single-particle dynamics in terms of the vibrational modes and show that our method yields a much more informative account of the trajectories' statistics.

\subsection{Contact vectors \textit{vs} contact forces}

$\{\vb{C}_i\}_{i=1}^N$ and $\{S_i\}_{i=1}^N$, defined respectively in Eqs.~\eqref{def:tot CV} and \eqref{def:tot dot CVs}, are predictive enough to allow us to identify a preferential direction of motion for each particle, as well as to spot the most mobile ones. However, the value of these observables do not depend on the magnitude of the contact forces but only on their direction. Although initially this might seem paradoxical, we can make sense of it from the following considerations: to begin with, from a purely static point of view and under the isostaticity condition, the forces magnitude can be obtained as the unique zero mode of the so called ``contact matrix'', $\mathcal{S}$, which can be constructed using only the contact vectors\cite{jamming_criticality_forces} (see Eq.~\ref{eq:S-matrix}), hence the magnitudes contain no extra information than what is already provided by the spheres position. Secondly, considering the MD dynamical protocol, it is clear that as soon as the spheres are no longer touching each other, the magnitude of a contact force loses its physical meaning, while the corresponding contact vector still contains information about which directions constrain a particle's motion the most.
Similarly, in the case of soft spheres the value of the contact force at jamming is not the relevant physical quantity either, because the real force driving the interactions of any pair of particles in the MC dynamics depends explicitly on both, their mutual overlap and the type of interaction potential, which we have parametrized here through $\nu$.
Nevertheless, independently of the specific value of the force magnitude the contact vectors can still be used as indicators of the optimal direction in which a particle should move in order to minimize the energy of the system because they contain information about the force field experienced by each particle, at least for the initial part of the dynamics. Additionally, as can be noted from the rightmost equality of Eq.~\eqref{def:tot dot CVs}, $S_i$ also incorporates the influence of the number of neighbours of each particle. This feature intuitively explains that independently of the vectorial contribution of the contacts, particles with more (less) neighbours are expected to be more (less) constrained. Besides it helps to understand why a significant correlation was also found when considering $\sum_{j \in \partial i} \abs{\vb{F}_{ij}}$ as a predictor, the reason being that this latter quantity is essentially a weighted sum of the number of contacts. Another fact to consider is that the distribution of dot products between contact forces and the associated one between contact vectors are significantly different, see Fig.~\ref{fig:dist-contacts} in Appendix \ref{sec:properties jamming}. This provides further evidence that these two quantities convey different information.

Even more importantly, the physics derived from the model of the fractal free energy landscape of amorphous solids gives another perspective to rationalize this feature. The argument goes as follows: in Ref.~\cite{fractal-free-energy} the authors measured numerically the overlap of contact \emph{vectors} between pairs of jammed states (\textit{i.e.}  pairs of minima in the landscape), defined by a target pressure $p^\star$, as a function of a common initial pressure of the configurations, $p_0$. What they found is that as $p_0$ was closer to $p^\star$, the overlap increased gradually. This shows that, although jammed configurations belonging to a same (fractal) basin share a similar structure, the specific neighbours with which a particle will eventually share a contact once in the jammed state are not defined unequivocally by the local environment of the particle above jamming, but instead they are determined gradually as this state is approached. By the same token it also indicates that as a configuration goes up in the landscape, the jammed state it departed from will still encode useful structural information, since a significant amount of the network of contacts is shared by the nearby minima. Now, note that we can think of the short time dynamics studied in this work as several realizations of trajectories followed by a configuration in its way to explore a small neighbourhood above its initial jammed state. Under these assumptions, the original basin should contain all the phase space available as this exploration takes place, which in turn implies that many of the interactions between nearest neighbours influencing the dynamics are going to remain essentially unchanged, and therefore we can expect major correlations with the actual network of contact vectors.

\subsection{Comparison with other methods and previous works} \label{sec:comparison-normal-modes}

As mentioned in the introduction, the dynamical properties of systems near their jammed state have been investigated predominantly in experiments, and for the scope of this work it is very relevant to consider Ref.~\cite{exper_coulais_how_2014}. In this study, the authors found a clear connection between the dynamics of their samples, which consisted in a bidisperse mixture of photoelastic soft disks, and the network of contacts formed among them. They studied both the under- and over-compressed parts of the dynamics, although an important difference with our work is that in their case the dynamics was driven by an external vibrating apparatus rather than by thermal fluctuations as we considered here. At any rate, our results should be considered as complimentary to each other because while we focused on the statistics of the trajectories for a given configuration, their results are based on several realizations on the network of contacts, \textit{i.e.} over several jammed states. The fact that during our simulations we found displacements of the same magnitude as the ones reported experimentally is reassuring.
This work is also important because their experiments managed to probe several of the dynamical features analysed theoretically by Ikeda \textit{et al.} in Ref.~\cite{dynamic-criticality-jamming}, specifically, the ones that the authors of this latter reference identify as a ``critical regime''. Such regime is characterized by strong anharmonicities that make a vibrational description unavailable, and as we will argue next, our simulations fall precisely within this same critical region. But before doing so we briefly mention the results obtained in this reference that are most important for our work. Their main finding is that the changes in the behaviour of the MSD can be associated with corresponding changes in the DOS, $D(\omega)$, where this latter quantity is calculated via the Fourier transform of the velocity autocorrelation function of the particles, $d(t)= \frac{1}{3N'T} \sum_{i=1}^{N'} \avg{\vb{v}_i(t) \vb{v}_i(0)}$ (as above, $N'$ is the number of non-rattlers). For instance, they showed that the time at which the MSD deviates from the ballistic regime because many particles begin to interact with their neighbours can be associated with a frequency for which the rapid decay of the DOS at large $\omega$ occurs. Similarly, the time at which the MSD reaches its asymptotic value, \textit{i.e.} the time it takes to reach the characteristic plateau present at high densities, has an associated frequency which indicates the onset of the Debye behaviour for small frequencies, $D(\omega)\sim \omega^{d-1}$. And although this identification is blurred out as the systems move away from their jamming point, say, by raising the temperature or changing the packing fraction, Ikeda \textit{et al.} were able to define a region where the harmonic description is always valid. Indeed, they showed that there is a temperature, $T^\star \sim \abs{\phi -\phi_J}^\nu$, such that whenever $T<T^\star$, then the collective properties of the dynamics are well described by the vibrational modes. In turn, if for a given $\phi$ the temperature of the system is large in comparison with $T^\star$, then the dynamics is dominated by anharmonic effects and no vibrational description is available and the system is in the so called critical regime.

Importantly, the instances considered in our work lie precisely in this latter regime as we will now argue. To begin with, recall that the MC simulations were performed exactly at $\phi=\phi_J$, meaning that for any finite temperature, the dynamics of the system will not be purely vibrational. In fact, the Ikeda \textit{et al.} state that in this scenario, the jamming DOS cannot be used to infer properties of the vibrational dynamics for any $T>0$. Additionally, the temperature range that we considered for the harmonic interaction, which is the only case analysed numerically by the authors, is several orders of magnitude above $T^\star$, so even if the harmonic description was appropriate at this packing fraction, all its effects in our simulations would have been washed out due to thermal noise.
The case of the MD results with hard spheres is more subtle, but as also pointed out in Ref.~\cite{exper_coulais_how_2014}, the experimental configurations analysed in \cite{exper_lechenault_critical_2008, exper_lechenault_super-diffusion_2010, exper_coulais_dynamics_2012}, which consisted of hard brass disks, also belong to the anharmonic regime. Clearly, these are empirical realizations that closely resemble the infinitely rigid spheres of our simulations, which suggests that our findings with $\phi<\phi_J$ should lay beyond an interpretation in terms of vibrational modes.
Yet, because we deem indispensable to provide a more rigorous verification that the normal modes do not furnish an adequate description of the single-particle dynamics in the regime we analysed here, we now proceed to present some results obtained using such method. We emphasize that our procedure cannot mimic exactly the one of Ref.~\cite{dynamic-criticality-jamming} because, as mentioned above, the authors used the velocity autocorrelation function in order to compute $D(\omega)$. However, note that $d(t)$ is intrinsically a dynamical property, while our approach relies on using exclusively knowledge about the structure of the configuration. This limitation is, fortunately, only apparent since using the formalism developed in Refs.~\cite{jamming_criticality_forces, lerner_low-energy_2013,degiuli_force_2014,lerner_breakdown_2014} we can compute the Hessian, $\mathcal{H}$, at the jamming point and then obtain the \emph{exact} DOS by diagonalizing it.
We remark that, formally, this is the procedure that should be followed to compute $D(\omega)$, although it has been shown\cite{pedagogical-vibrational-modes} that, under certain conditions, it coincides with the DOS obtained from the Fourier transform of $d(t)$, which is the technique most commonly utilized. Following Ref.~\cite{jamming_criticality_forces}, once we have the $N_c$ contact vectors $\vb{n}_{ij}$ introduced in Sec.~\ref{sec:statics}, we can compute the $N_c\times d N'$ contact matrix $\mathcal{S}$ whose components are given by
\begin{equation} \label{eq:S-matrix}
	\mathcal{S}_{(ij)}^{k\alpha} = \qty(\delta_{jk} - \delta_{ik})n_{ij}^\alpha
\end{equation}
where, as before, $(ij)$ denotes an ordered pair, and $\alpha=1,\dots,d$ is an index running through the dimensionality of the system. As a side note, we stress the fact that the vibrational formulation is unable to take the rattlers into account, since they are omitted from $\mathcal{S}$ by construction; in contrast, our method incorporates them readily.

In the case of an harmonic contact potential, the Hessian at the jamming point is given by $\mathcal{H}=\mathcal{S}^T \mathcal{S}$, but universality ensures that a jammed state with a given potential is also a valid jammed configuration for any other potential\cite{jamming_criticality_forces,parisi-et-al-jamming-annrev}. In the case of hard spheres, Eq.~\ref{eq:S-matrix} should be multiplied by magnitude of the contact forces, \textit{i.e.} $n_{ij}^\alpha \to f_{ij} n_{ij}^\alpha$, thus obtaining the following expression for the Hessian:
\begin{equation}\label{eq:H-hard-spheres}
\mathcal{H}_{i\alpha}^{j\alpha'} = \qty(\widetilde{\mathcal{S}}^T \widetilde{\mathcal{S}})_{i\alpha}^{j\alpha'} =
 \delta_{ij} \sum_{k\in \partial i} f_{ki}^2 n_{ki}^\alpha n_{ki}^{\alpha'}
- \delta \qty[(ij)] f_{ij}^2 n_{ij}^\alpha n_{ij}^{\alpha'}
\end{equation}
where $\widetilde{\mathcal{S}}$ is the modified contact matrix taking into account the forces magnitude, and $\delta\qty[(ij)]$ is a function that is one only if particles $i$ and $j$ are in contact and zero otherwise. To recover the expression for the harmonic interaction it suffices to omit the factors containing the forces magnitudes.

To obtain the normal modes of the system, we simply need to obtain the eigenvalues of the Hessian, which in turn determine the normal frequencies of the system and, thus, the DOS: if $\{\lambda_q\}_{q=1}^{3N'}$ are the eigenvalues of $\mathcal{H}$, then the associated frequency is $\omega_q=\sqrt{\lambda_q}$. The $3N'$ eigenvectors\footnote{We will use the notation $\va{\bullet}$ to denote the $3N'$ dimensional vectors of the configuration space.}, $\{\va{v}_q\}_{q=1}^{3N'}$, correspond precisely to the vibrational modes, each of which describe a unique oscillatory mode of the configuration around the jamming energy minimum. For further use, we introduce here the Inverse Participation Ratio (IPR), which can be calculated in terms of the eigenvectors as,
\begin{equation}\label{eq:IPR}
	Y(\omega) = \frac{\sum_{i=1}^{N'} \abs{\vb{v}_i(\omega) }^4  }{\qty(\sum_{i=1}^{N'} \abs{\vb{v}_i(\omega)}^2 )^2}
\end{equation}
where $\vb{v}_i(\omega)$ is a $d=3$ dimensional vector obtained from the eigenvector $\va{v}$ with corresponding frequency $\omega$ by taking the components associated with particle $i$. Intuitively, $Y(\omega)$ quantifies the localization of the mode with frequency $\omega$; that is, for very small values of the IPR, $Y\sim 1/N'$, the associated mode is essentially extended, implying that many particles are collectively displaced, all of them by roughly the same amount; while for large values, \textit{i.e.} $Y\sim 1$, few particles participate in the displacement excitation, so the mode is localized.

\begin{figure}[htb!]
	\includegraphics[width=\linewidth]{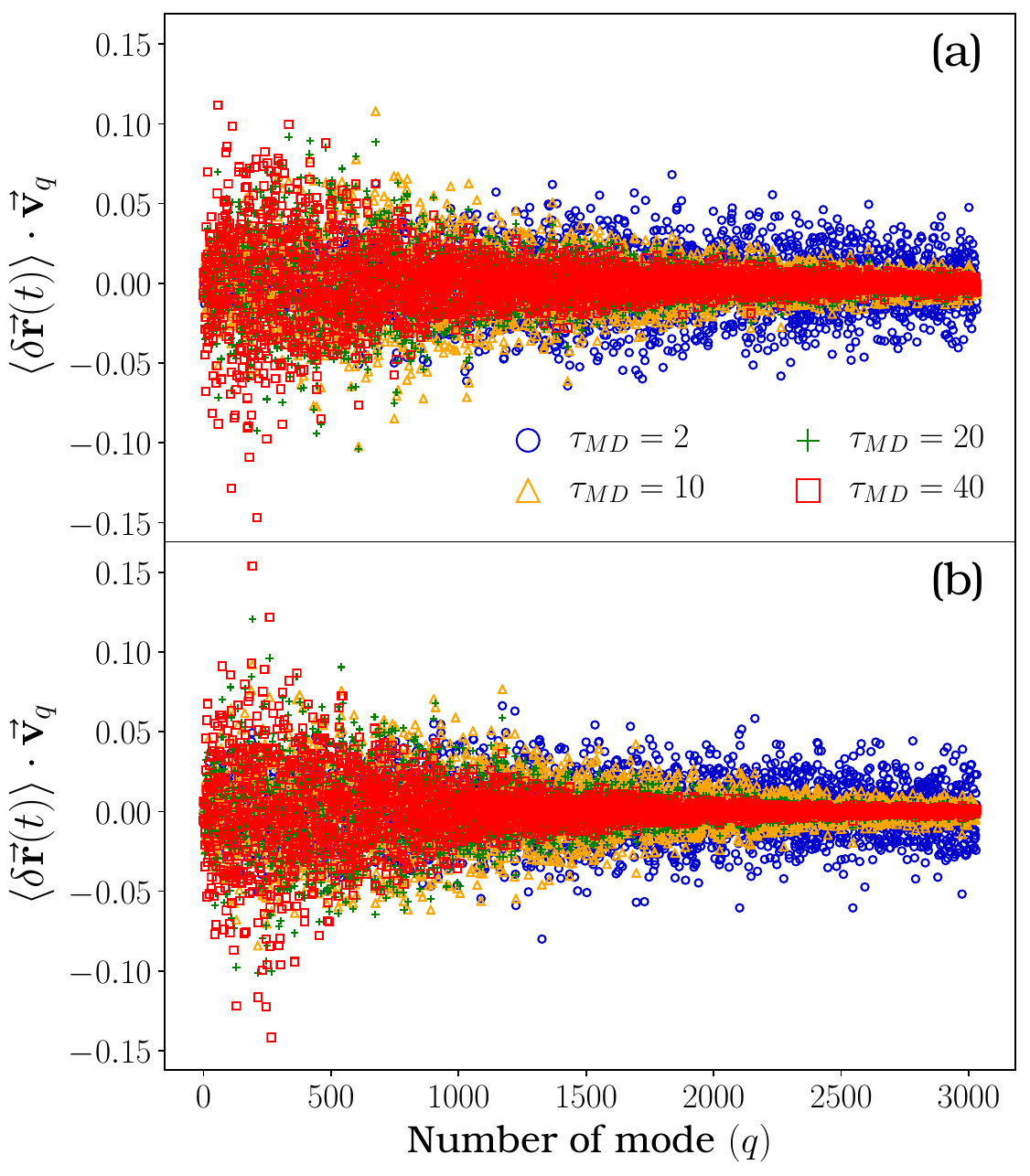}
	\caption{Value of the projection of $\avg{ \delta \va{r}(t)}$ along each of the normal modes, $\{\va{v}_q\}_{q=1}^{N'}$, obtained from the Hessian using the harmonic (panel a) or hard-sphere interaction potential (panel b). The different times at which the dot products were computed are identified by different colours and markers, according to the legend. Values of $\avg{ \delta \va{r}(t)}$ were obtained from the simulations with $\phi/\phi_J=0.995$. It is only for later times that more weight can be assigned to few of the low frequency modes, but there is no way of knowing, \textit{a priori}, how to select them.}
	\label{fig:projections-mean-disp-per-mode}
\end{figure}

\begin{figure}[htb!]
	\includegraphics[width=\linewidth]{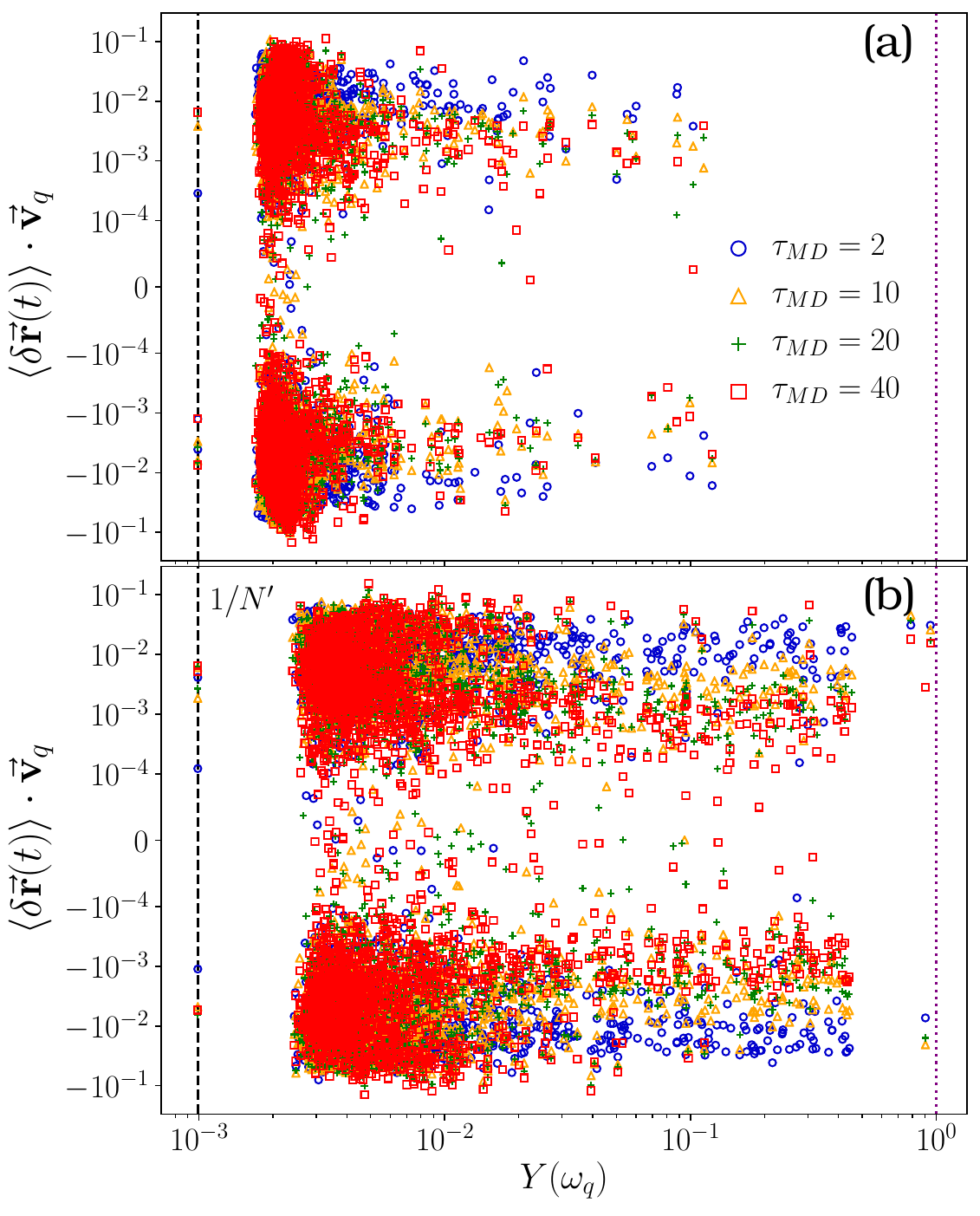}
	\caption{Same projections as in Fig.~\ref{fig:projections-mean-disp-per-mode} but plotted as a function the IPR defined in Eq.~\eqref{eq:IPR}, using the harmonic and hard-sphere potentials, panels (a) and (b) respectively. The black vertical dashed line at $1/N'$ is the lowest value that can be attained by $Y(\omega)$ and correspond to maximally extended modes, while the dotted line at $1$ identifies the most localized ones. The few points with $Y(\omega)\approx 1/N'$ are associated with the $d=3$ zero modes caused by the translational invariance of our systems and their respective weight can be used as a lower threshold for determining when a mode can be considered relevant (see main  text).}
	\label{fig:IPR-vs-projections}
\end{figure}

Now, it is clear that if the particles were vibrating around the energy minimum their displacements would be easily captured by the normal modes formalism. Moreover, it should be expected that if this is the case, then the lowest energy modes should be the ones carrying most of the weight of the configuration's displacement. To put to test this assumption, we measured the projections of the configurational vector of mean displacement, $\avg{ \delta \va{r}}$, along the different modes. The resulting values from the MD simulations with $\phi/\phi_J=0.995$ are reported in Fig.~\ref{fig:projections-mean-disp-per-mode}, where the Hessian was obtained with both the harmonic (upper panel) and hard sphere (lower panel) interaction potentials, while different colours correspond to different times at which $\avg{ \delta \va{r}}$ was calculated. We mention in passing that all the results presented in this section were obtained after $\avg{ \delta \va{r}}(t)$ was normalized at each time, so that the different values of the projections remain comparable to each other as time evolves. The scattering of the points evinces that, for either type of potential, there is no clear signature that few modes are responsible for the collective particles' displacement, specially for the initial times. For longer times, more weight is concentrated in the first modes, but still there are many other low frequency modes with negligible weight. We are of the mind that such lack of criterion to select the most relevant modes exhibits one important drawback of the normal mode description at the single-particle level, and, unfortunately, it also affects other properties such as the IPR.
This is important because, naturally, we could suppose that not all the modes are going to be equally important, independently of their frequency, due to the fact that the trajectories we studied here consist in displacements of all the particles of the system. This suggests that what we should be actually investigating  is whether the extended modes posses a higher weight than the localized ones. Were this to be the case, then choosing the modes with the lowest value of $Y(\omega)$ could be taken as a criterion for selecting the vibrational modes that influence the dynamics the most. Yet, as Fig.~\ref{fig:IPR-vs-projections} demonstrates, also when using the IPR we are unable to identify the correct normal modes to account for the configuration's displacement statistics, since even when the IPR is very small, \textit{i.e.} $Y(\omega) \in (2\times 10^{-3}, 10^{-2})$, the corresponding values of the projections lie within a range of 2-3 orders of magnitude. Such a broad distribution exemplifies that the collective dynamics cannot be described as if it consisted of few extended modes. Note that, on the one hand, these results are independent of the type of potential, while on the other, allowing the dynamics to evolve for longer times only affects the localized modes, the vast majority of which loses most of its weight.
A more quantitative comparison is at hand if we consider the weights of the three most extended modes (the ones lying on the vertical line $1/N'$) corresponding to the zero modes of $\mathcal{H}$, which in turn are associated with the translational invariance of our system induced by its periodic boundary conditions. These modes determine directions along which the system could be collectively displaced without any energy cost, but since we know that during the dynamics the systems is \emph{not} being uniformly translated, we can use their weight as a lower threshold for when a given mode is relevant for the dynamics. What we observe from Fig.~\ref{fig:IPR-vs-projections} is that there are several modes (also considerably extended) that have an even smaller weight, while we still lack a guideline to find the ones more aligned with $\avg{ \delta \va{r}}$. Furthermore, note that even the sign of the projection is important if we really want to be able to identify preferential directions, but the vibrational scheme fails to determine such information because it is assumed that the motion will consist mainly of oscillations around a given state, so the direction of motion becomes irrelevant. Nonetheless, our results of the previous sections prove that this is not the case for the dynamical regime we are studying. For completeness, we mention that we also verified that our results are unchanged if instead of the value of the dot product we had used $K\qty[\avg{ \delta \va{r}}, \va{v}_q]$ for measuring the correlation between the configurational displacement and the normal modes.

\begin{figure}
	\includegraphics[width=\linewidth]{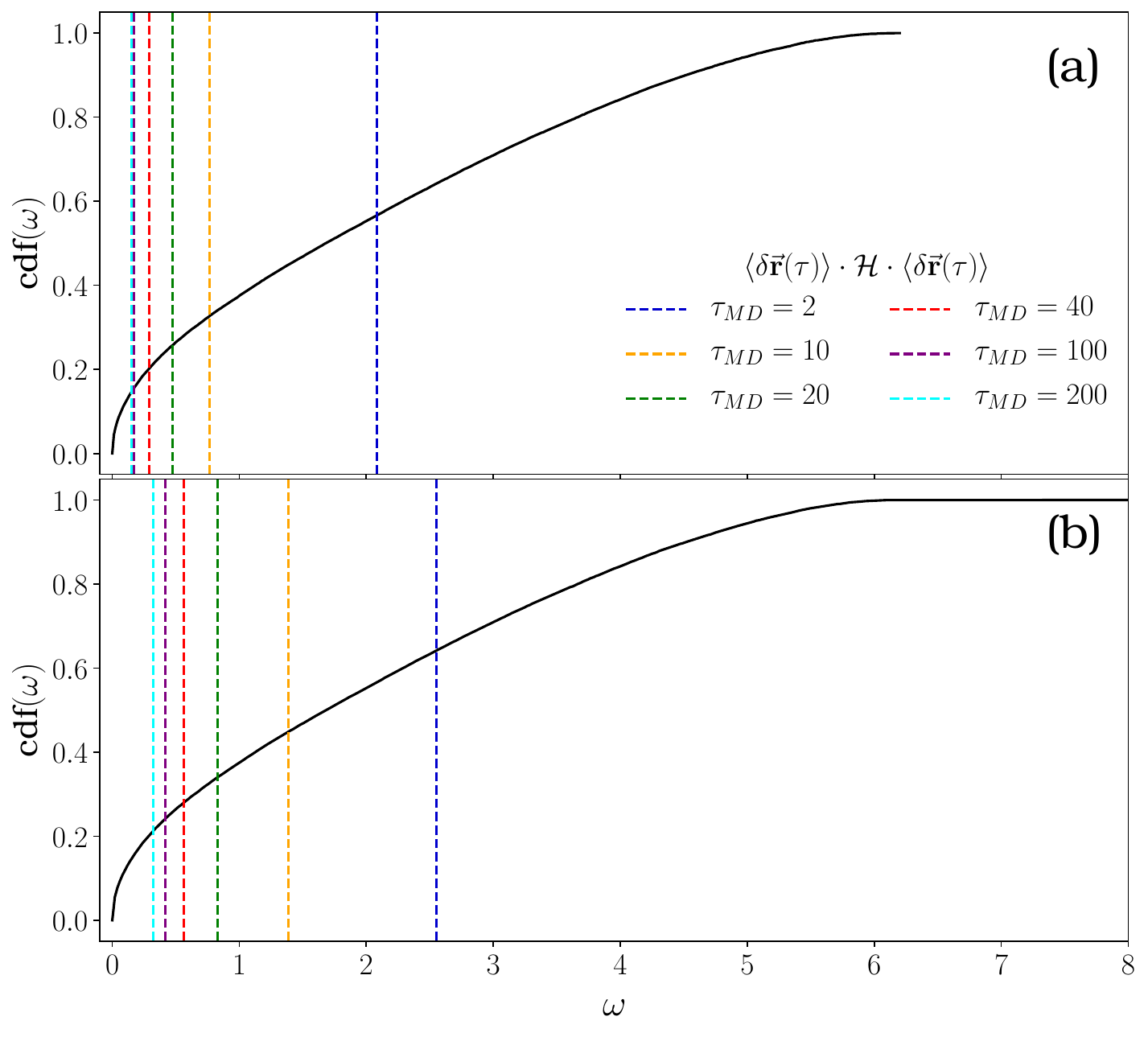}
 	\caption{Comparison of the ``curvature'', measured by $\avg{\delta \va{r}(t)}\cdot \mathcal{H} \cdot \avg{\delta \va{r}(t)}$, with the spectrum of $\mathcal{H}$ for the harmonic (top) and hard-sphere (bottom) interaction types. Values of this curvature are indicated at several times by the vertical dashed lines, and since the configurational mean displacement is normalised at each time, their intersection with cdf$(\omega)$ (black curve) equals the fraction of frequencies smaller than such value. The fact that the fraction of frequencies smaller than the value of curvature is always extensive, even for long times, reveals that the dynamics we probed during our simulations are not described by the forces that could be derived following the vibrational approach.}
	\label{fig:curvature-hessian-cdf-omega}
 \end{figure}

Such lack of connection between the mean configurational displacement and the distinct vibrational modes can be explained, as we argued before, by the fact that these modes are not suitable to examine the trajectories generated with our simulations because they belong to the critical region where the harmonic approximation is not valid. And even though the results presented so far support this hypothesis, we have not yet verified that the dynamics is mainly driven by anharmonic effects. To do so, we computed the value of $\avg{\delta \va{r}(t)}\cdot \mathcal{H} \cdot \avg{\delta \va{r}(t)}$, which should provide an estimate of the ``curvature'' at the minimum induced by the dynamics. More specifically, if the harmonic description was valid, this curvature should be small because that would indicate that the particles are displaced along the lower-energy (\textit{i.e.} flatter) directions. Conversely, a large value signals that the forces driving the dynamics are not captured by the linear assumption implicit in the harmonic assumption. To put both of these scenarios to test, in the upper (resp. lower) panel of Fig.~\ref{fig:curvature-hessian-cdf-omega} we plot the cumulative distribution function (cdf) of the frequencies for the harmonic (resp. hard-sphere) potential and we indicate the values of $\avg{\delta \va{r}(t)}\cdot \mathcal{H} \cdot \avg{\delta \va{r}(t)}$ with vertical dashed lines, using different colours to distinguish different times according to the legend. Recall that since we have normalized $\avg{\delta \va{r}(t)}$ at each time, the intersection of each vertical line with the curve of cdf$(\omega)$ determines the fraction of frequencies smaller than the corresponding value of the curvature. Clearly, during the initial times the value of the curvature is quite large, \textit{i.e.} greater than $40-60\%$ of the frequencies, specially considering the hard-sphere potential. But, importantly, even for longer times this fraction never gets much smaller than the $20\%$, which is still an extensive number of modes, thus corroborating that the preferential directions of motion, inferred via the vibrational modes of low energy, do not correspond to the real trajectories followed by the particles. This finding establishes that our systems were in the anharmonic region, according to the classification of Ref.~\cite{dynamic-criticality-jamming}.


\section{Conclusion and perspectives}\label{sec:conclusions}

We presented a simple and robust method to connect the structural information present at the jamming point of a configuration of $N$ spherical particles and the dynamics that happens close to such point. We first analysed the statistics of the spheres' displacement after producing several trajectories originating from the positions at the jammed state. The resulting distributions show, on the one hand, the existence of preferential directions of motion, and on the other, the presence of heterogeneity in the particles' mobility, reflected by the fact that some spheres remain mostly fixed, while others are more mobile by several orders of magnitude, see Fig.~\ref{fig:pdf displacement}. We also argued that these features are captured by the respective first moments, \textit{i.e.} $\avg{\delta \vb{r}_i}$ and $\avg{\abs{\delta \vb{r}_i}^2}$ for $i=1,\dots,N$, and thus we used these statistics to succinctly represent the full distributions.
Our main result is that we were able to find a straightforward and significant connection, at the single particle level, between these statistics and the network of contact vectors that is formed at the jammed state. In particular, we found that the sum of contact vectors acting on a given particle, $\vb{C}_i$ defined in Eq.~\eqref{def:tot CV}, is a good predictor of its preferred displacement direction, if any, while the sum of dot products between all pairs of contact vectors,  $S_i$ defined in Eq.~\eqref{def:tot dot CVs}, is highly correlated with the particle's mobility. We tested our approach using two different types of dynamical protocols, namely Molecular Dynamics and Monte Carlo simulations, and verified that it remains valid in both cases and for the different parameters we studied; cf. Figs.~\ref{fig:MD-displacements-vs-totCV}-\ref{fig:ranking-forces-and-displacements}.
The statistical correlation, measured by the Spearman coefficient, $K$, decays rather slowly as the dynamics evolves, thus showing that the jammed state used as the initial condition of the trajectories has a persistent influence on the particles displacements. More explicitly, using the MSD as a measure of how far the configuration is from its initial state and comparing the value of correlation as a function of $\Delta$, we found that while $K$ decreases its value by an order of magnitude or less the MSD has increased by at least two decades. Because we are not probing whatsoever the ballistic regime, such change in the MSD is only due to interaction between particles, and thus implies that $\vb{C}_i$ and $S_i$ are relevant even after a small but clearly measurable rearrangement of the whole configuration has taken place. Furthermore, in Fig.~\ref{fig:correlations-vs-t} we showed that when the value of $\Delta$ is rescaled by a factor that depends on how far from the jamming point the dynamics occurs, the curves of the loss of correlation can be superimposed to each other. This suggests that for a given dynamical protocol the decorrelation rate is, surprisingly, independent of the distinct parameters used in our simulations. We therefore conclude that the network of contact vectors formed at a jamming point contains significant information about the short-time features of the particles' motion near such point.

Importantly, our results show that in the dynamical regime we studied, and for the timescales of our simulations, the particles' motion does not consist of simple vibrations around a reference equilibrium state, which in our case is naturally identified with the initial jammed configuration. This novel finding is supported by the fact that the vibrational modes fail to capture the real mean displacement of the configuration, as evinced by the negligible correlations we reported in Figs.~\ref{fig:projections-mean-disp-per-mode} and \ref{fig:IPR-vs-projections} of the previous section. We emphasize that the scope of this work is restricted finding a connection between dynamical and structural (and hence, static) variables, therefore we only used data about the contact forces as input for our statistical inference. These same data can be used to formally compute the Hessian at the jamming point, whence the normal modes can be easily obtained. This shows that, utilising the same structural information as input, our approach yields a better description of the statistics of the single-particle dynamics compared to previous works and obtain comparable correlations compared with more sophisticated, state of the art methods.
Additionally, we want to stress that because our method is rather simple and straightforward to apply, we are optimistic in that it can be used to investigate other type of systems that share jamming as a common critical point. In other words, given that the same critical phenomenology can be reached by tuning different physical variables it is likely that the structural variables we identified here, namely $\vb{C}_i$ and $S_i$, should also convey information about the particles trajectories when the dynamics is driven by, say, the application of a load or a stress. Moreover, jamming criticality has recently been shown to be very robust with respect of dimensionality and type of interaction potential, so our approach presumably has the same range of validity. On the other hand, let us recall that the jamming universality class is composed not only by amorphous solids and glass formers but also by several constraint satisfaction problems and learning algorithms, where active research has been carried out to explore how their corresponding jammed states are reached. Thus, once the analogous structural variables are identified in those cases, our method could also be used to extract new insights about the dynamics of the algorithms utilized in that sort of problems.

Also necessary is a word about the regime of validity of our method, for which we will once again adopt the picture of the fractal free energy landscape, in particular the results of Ref.~\cite{fractal-free-energy}. There it was shown that jammed state is realized gradually as the configuration goes down towards a minimum and the contact forces are determined. The authors verified that similar jammed states exhibit similar network of forces and, more importantly, different realizations of jammed states, originated from a single initial configuration, would have a significant fraction of common contacts; of course, the closer this initial configuration is to an energy minimum the higher the fraction of common contacts in the final jammed packings. Now, assuming that the short time dynamics essentially explores a small vicinity around the reference jammed sate, these findings set the Gardner phase as a rough limit beyond which we do not expect our method to remain valid. That is as long as the dynamics departs from a state (in parameter space) within this marginal phase, it is likely that we find non-negligible correlations between our structural variables and the single-particle displacement statistics.
To see why this is the case, let us imagine that we use a jammed configuration as the initial condition of a given dynamical protocol. After a short time we stop the evolution, then bring the configuration to a new jammed state and, finally, we restart the dynamics using this second state as initial condition. Since the two jammed states are similar, we would expect that the second trajectory to be correlated, at least to some extent, with the structure of the first jammed state. Yet, note that for this scenario to be true it is crucial that the dynamics occurs inside the Gardner phase because only in this way we can guarantee that the structural variables from the original configuration are going to be similar to the ones of the nearby basins explored during the dynamics. Nevertheless, this argument does not provide a criterion for determining the rate and sharpness of the loss of correlation as the stable glass phase is approached. This and other open questions are describe in more detail next.

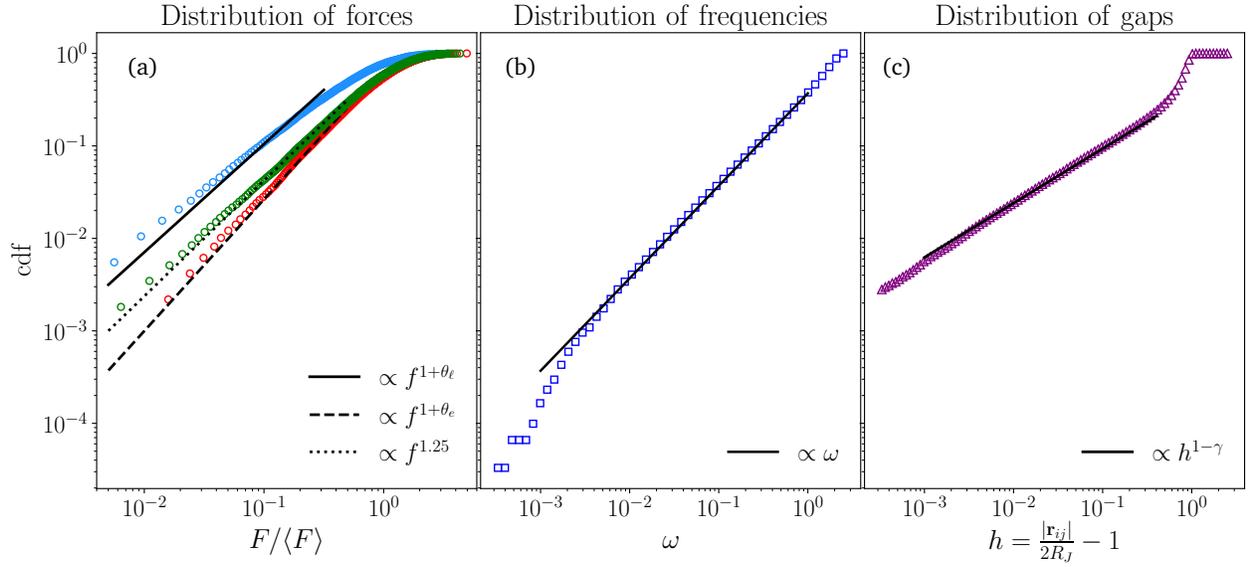
\begin{figure*}[!htb]
	\centering
	\setlength{\unitlength}{0.1\textwidth}
	\begin{picture}(10,4.3)
	\put(0.4,0){\includegraphics[width=0.9\textwidth]{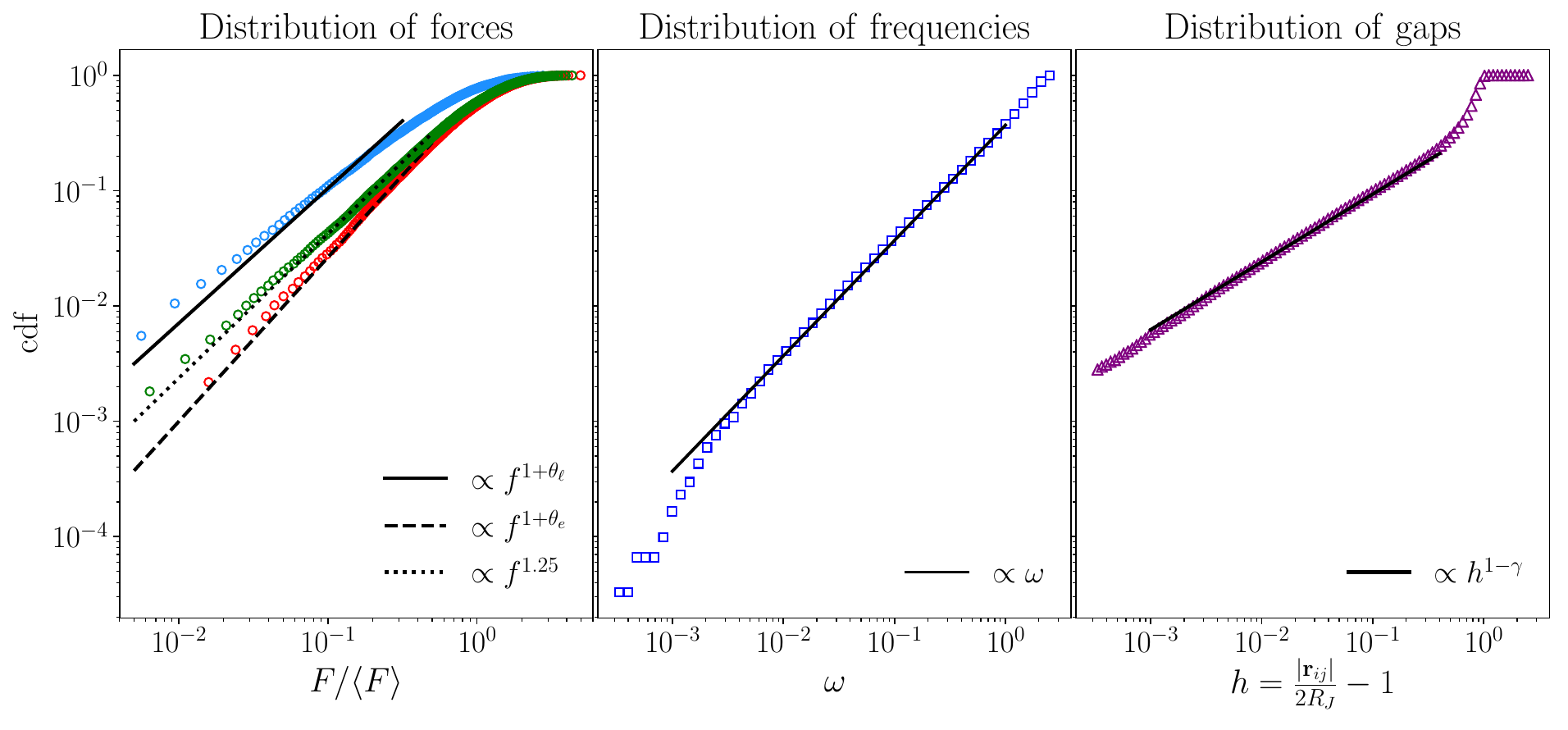}}
	\put(1.3, 3.6){(a)} \put(4.0, 3.6){(b)} \put(6.7, 3.6){(c)}
	\end{picture}
	\caption{Cumulative distribution functions (cdf) of contact forces (panel a), frequencies (panel b), and interparticle gaps (panel c). In panel a, we have split the forces in the ones associated to bucklers (blue) and extended modes (red), following Ref.~\cite{jamming_criticality_forces}; the black lines are comparisons with the power laws predicted by the theory. The green markers are all the forces put together, while the power of the associated fit was taken from the same reference. In panel b we used the fact that the density of states is predicted to be constant for small values of $\omega$ and hence its cdf should be proportional to the frequency as verified by comparing with the black line. Finally, in panel c we show the cdf of the gaps ($h$, defined in the figure's horizontal axis) between nearby particles; theory establishes that the probability density scales as $h^{-\gamma}$ for small gaps and our results are in excellent agreement with such a prediction.
	}
	\label{fig:properties-jamming}
\end{figure*}

We finish by pointing out several routes of improvement that can be considered. The first one is to construct other variables that incorporate information of the ``second nearest neighbours''. In other words, as it is now, our method only considers the role of a particles first neighbours, but it is expected that a more complete description can be attained when a larger vicinity is considered. An independent issue to consider is that with enough data, we could in principle aim to infer the full distribution of the displacements of particles instead of just their first moments as we did here. This would provide a complete description of the trajectories followed by amorphous solids near their jamming point. It is clear that both approaches can be combined, where the degree of statistical agreement of the full distribution is increased by taking into account more complicated model, which in turn would include observables constructed from an even larger set of particles neighbours and more parameters. This opens the way to the application of Bayesian model selection which has proven to be very fruitful in a surprisingly broad range of topics.
Along the same lines, it is also important to notice that we used a particle-wise approach in which the most relevant structural features are captured by the position of a particle's nearest neighbours. But one could also think on implementing a more ``coarse-grained'' description in which the dynamics of a group of particles (say their mean mobility) is characterized by a structural quantity of the corresponding region (for instance the average value of $S_i$ in the cluster). This clusterized approach would be similar to the more common methods that explore the influence of the local environment on the dynamics of glassy systems, hence we expect that the more we adopt a cluster description the longer the times our description should remain valid, as reported in other works. In this way, one could extend the regime of high correlations by including the information of more neighbours and analysing their statistical properties.
The possibilities just mentioned are concerned with developing techniques to improve the quality and duration of the statistical inference, but our results also hint at some properties that deserve to be explored, from a physical point of view, with greater detail. To begin with, given that previously it has been shown that particles are dynamically correlated at the large scale, while we have shown that there are also important local structural correlations, it is to be expected that $\vb{C}_i$ and $S_i$ (or other similar structural variables) should also exhibit signatures of correlations on longer length scales. Unfortunately, due to high fluctuations in these structural variables a very large number of independent jammed configurations should be used in order to extract a meaningful signal and thus identify the presence of truly extensive structural correlations. Nonetheless, we deem that putting this hypothesis to test is particularly relevant since it would provide stronger evidence of the influence of the local structure on the dynamics.
Another promising path for extension, also related to long range correlations, is how the presence of system-wide order or regularity would influence our results. Candidates for such studies could be, for instance, either low density systems, such as tunneled crystals\cite{jiao_nonuniversality_2011} or, on the contrary, high density ones, formed by slightly polydisperse crystals\cite{gardner_physics_crytals, tsekenis_jamming_crystals}. Besides the usual marginal stability of common jammed packings, which leads to structural correlations lengths comparable to the system size, these systems are characterized by periodic structural order of their particles; therefore, two different types of structural correlations coexist across the whole configurations. It is clear that our method can also be directly applied to such systems because the network of contacts is also well defined in all of them, but the extent to which the inherent order of these configuration increases the relevance of the second nearest neighbours and influences the dynamics, and thus the level of correlation our approach would yield, is something that remains to be explored.
Another independent feature to investigate is how the dynamics of the configuration in its original basin affects the loss of correlation in time. In other words, it would be interesting to study if there is a connection between this decorrelation and the distance from the original minimum to nearest one to the state of the configuration.
This would provide evidence that the decorrelation is strongly influenced by the overlap between the many jammed sates in a basin of the free energy landscape and, moreover, it could suggest how to use the structure of these other free energy minima to enhance the quality and the duration of the statistical inferences we presented here or other similar ones. Additionally, if a more direct link between such statistical correlations and similarity between the jammed sates within a basin is found, it could be used to set a more accurate limit for the range of applicability of the sort of inference tools we developed here.
However, we leave these unexplored areas as topics for future research.

\section*{Conflicts of interest}
There are no conflicts to declare

\section*{Acknowledgements}
The research has been supported by the European Research Council under the European Unions Horizon 2020 research and innovation programme (grant No. 694925, G. Parisi) and the project “Meccanica statistica e complessità”, a research grant funded by PRIN 2015 (Agreement No. 2015K7KK8L). We are grateful to Patrick Charbonneau and Georgio Tsekenis for valuable suggestions and to David Becerril proof reading our manuscript. We also thank the anonymous referees for their insightful comments that helped us enhance the presentation and significance of our work. RDHR thanks Angelo Cavaliere for useful discussions during the preparation of this work.

\appendix

\section{Details of the Linear Programming (LP) algorithm} \label{sec:details algorithm}


LP-type algorithms have been used before to produce jammed configurations of hard spheres, see for instance \cite{torquato_robust_algorithm,donev_linear_2004}, so we will only briefly describe the implementation of our method. In the referred works, the volume containing the spheres is modified in a sequence of steps until the jamming point is reached and the way in which the volume is to be modified in each step is determined by the optimal solution of a LP problem. Recently, in \cite{artiaco_baldan_parisi} the authors adopted a similar approach with the difference that they kept the system's volume fixed (namely, a cubic box of size $L\sim N^{1/3}$ and periodic boundary conditions) and instead increased the particles' radius at each step. We essentially followed this latter version, but we also made explicit use of the dual variables associated to the non-overlapping constraints in order to access the contact forces between particles as we explain now. We began with a low density configuration (\textit{i.e.} $\phi < \phi_J$) where the particles' position and size are, respectively $\vb{r}_i$ and $R< R_J$. Producing a jammed configuration (although not necessarily isostatic) is equivalent to finding the displacements, $\{\vb{s}_i\}_{i=1}^N$, that allow to obtain a new configuration, where each of the particles position is updated according to $\vb{r}_i \to \vb{r}_i + \vb{s}_i$,  and their size can be maximally increased, say by a factor $\sqrt{\alpha}$, but with no overlap whatsoever between any pair of them. In other words, we can recast the problem of producing a jammed configuration as the following optimization problem
%
\begin{equation}\begin{aligned} \label{def:otimization problem}
\text{maximise } & \alpha \quad \\
\text{s.t.} \quad || \vb{r}_i + \vb{s}_i - (\vb{r}_j + \vb{s}_j) ||^2 & \geq \alpha R^2 \qc
\quad \forall\ i<j = 1,\dots, N \, .
\end{aligned} \end{equation}
%
Unfortunately, this is a non-convex optimization problem and thus there is no general method to solve it. Nonetheless, we can make progress by assuming that the displacements vectors are small and thus, by neglecting terms involving products of the type $\vb{s}_i \cdot \vb{s}_j$ the non-overlapping constraints for all pair of particles are transformed into a set of linear constraints. Because the objective function is also linear, the original problem is reduced to an LP one:
%
\begin{equation}\begin{aligned}\label{def:LP problem}
\text{maximise } & \alpha \\
 \text{s.t.} \quad || \vb{r}_i - \vb{r}_j||^2 + 2 (\vb{r}_i - \vb{r}_j)\cdot (\vb{s}_i - \vb{s}_j) & \geq \alpha R^2 \qc\\
\qquad \forall\ i<j = 1,\dots, N \ .
\end{aligned}\end{equation}
Note that the neglected term that lead to \eqref{def:LP problem} is positive, thus, any solution to this linear problem will also be a solution to the original exact problem, \eqref{def:otimization problem}. Hence, our algorithm allows us to exactly maintain  the hard-sphere constraint for all times.

In practice however, the optimal solution of the linearized problem does not produce a jammed configuration (in terms of the exact constraints) and a sequence of LP optimizations should be carried out in the following way: $N$ random vectors are drawn uniformly, $\{\vb{r}_i^{(0) } \}_{i=1}^N$, and an initial size $R^{(0)}$ is chosen (\textit{e.g.} as half the distance between the closest pair of points). With these values we solved the associated LP problem described above and then utilized such optimizers to generate a new LP problem whose associated variables are, naturally, $\vb{r}_i^{(1)} = \vb{r}_i^{(0)} + \vb{s}_i^{(0)}$ and $R^{(1)} = \sqrt{\alpha^{(0)}} R^{(0)}$. We repeated this process until convergence was reached, which is characterized by $\alpha=1$ as the optimal solution of the LP problem. All the optimization steps of this sequence were carried out using the Gurobi Solver\cite{gurobi} with the JuMP\cite{jump} interface.

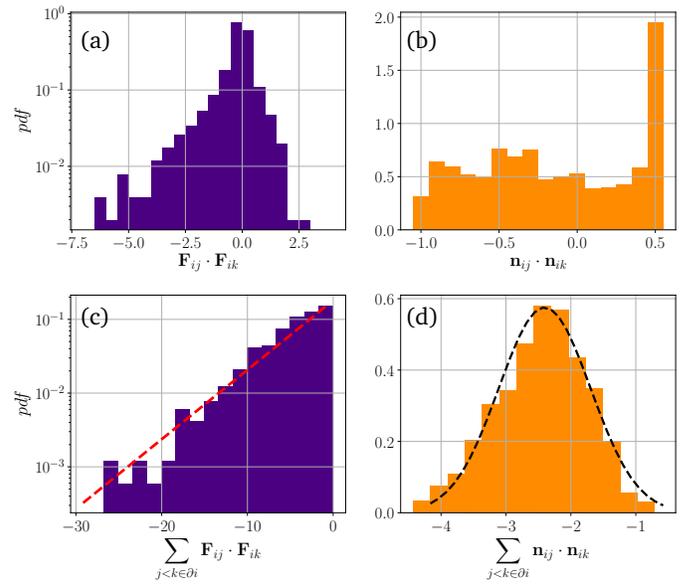
\begin{figure}[!htb]
	\centering
	\setlength{\unitlength}{0.1\linewidth}
	\begin{picture}(10,9)
	\put(0.0,0){\includegraphics[width=1.0\linewidth]{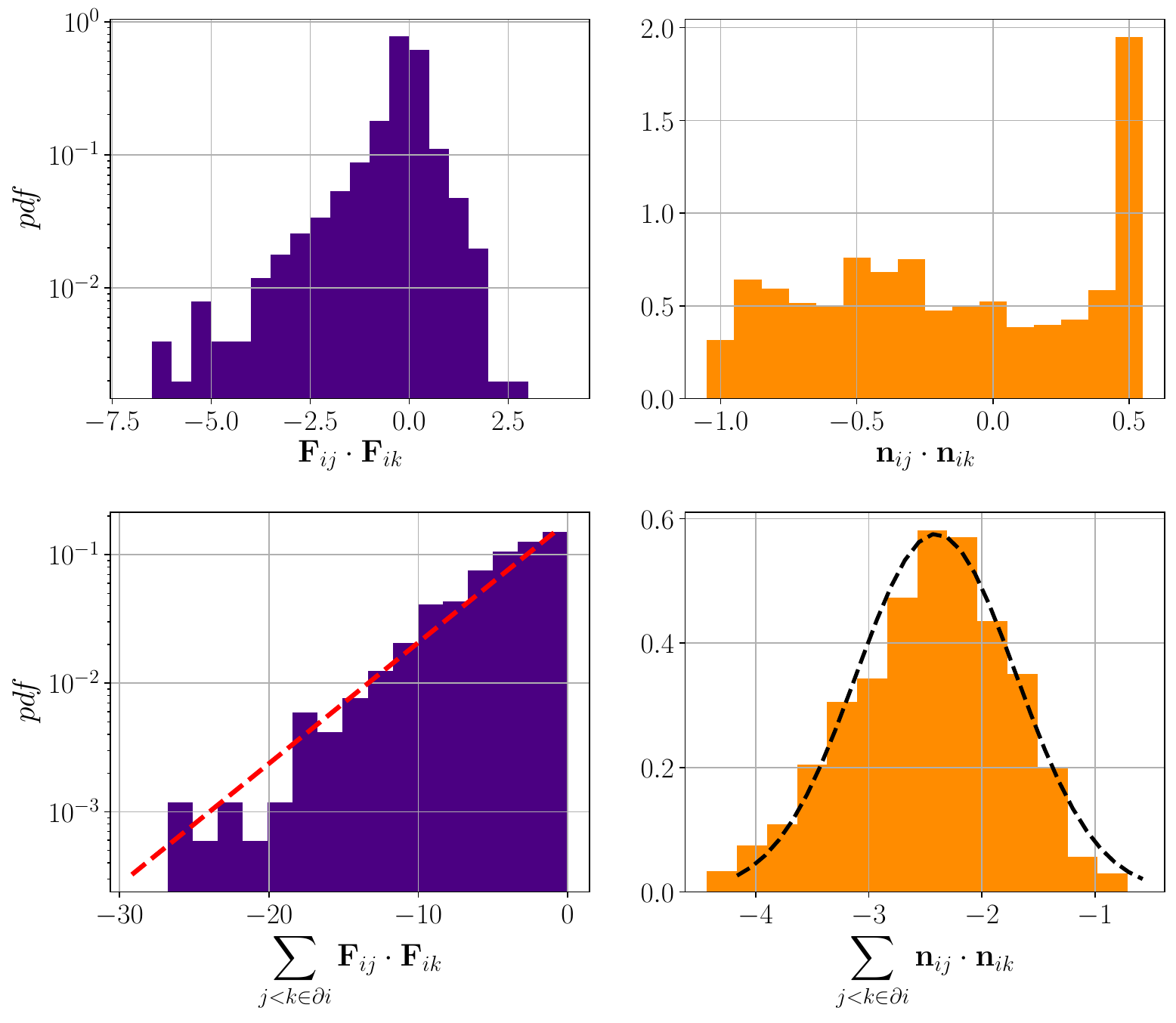}}
	\put(1.07,8.0){(a)} \put(5.9,8.0){(b)}
	\put(1.07,3.9){(c)} \put(5.9,3.9){(d)}
	\end{picture}
	\caption{Distribution of dot products of all forces (upper left panel) and contacts pairs (upper right) acting on the particles and their sum (lower panels). The data used in this figure was taken from the configuration used for the MC and MD simulation of this work, excluding rattlers.}
	\label{fig:dist-contacts}
\end{figure}

Importantly, once the jamming point has been reached, by computing the dual variables associated to the non-overlapping constraints of the problem \eqref{def:LP problem}, we can access the contact forces between particles whereby we can construct the force-network of the jammed configuration as shown in the right panel of Fig.~\ref{fig:jammed configuration}.

\section{Properties of the jammed configuration} \label{sec:properties jamming}

With the algorithm just described we were able to produce isostatic jammed configurations that, as anticipated in the main text, exhibit the same behaviour as the systems generated with other methods as well as fulfil the scaling properties predicted by the theory. Particularly, in Fig.~\ref{fig:properties-jamming} we show the distribution of forces (a), the density of states (b), and the distribution of gaps between particles (c), obtained from 10 typical jammed configurations of $N=1024$ particles produced via our LP algorithm just described, and compare them with the theoretical scalings. In the first case, the forces distribution is characterized by two exponents, $\theta_\ell \approx 0.17$ and $\theta_e \approx 0.42311$, while in the case of the particles gaps the relevant one is $\gamma \approx  0.41$. The configurations used in this work, either to perform the simulations reported in the main text or the ones reported in the Appendix \ref{sec:second-configuration}, are two independent systems selected at random from this set.

\begin{figure}[!htb]
	\centering
	\includegraphics[width=0.99\linewidth]{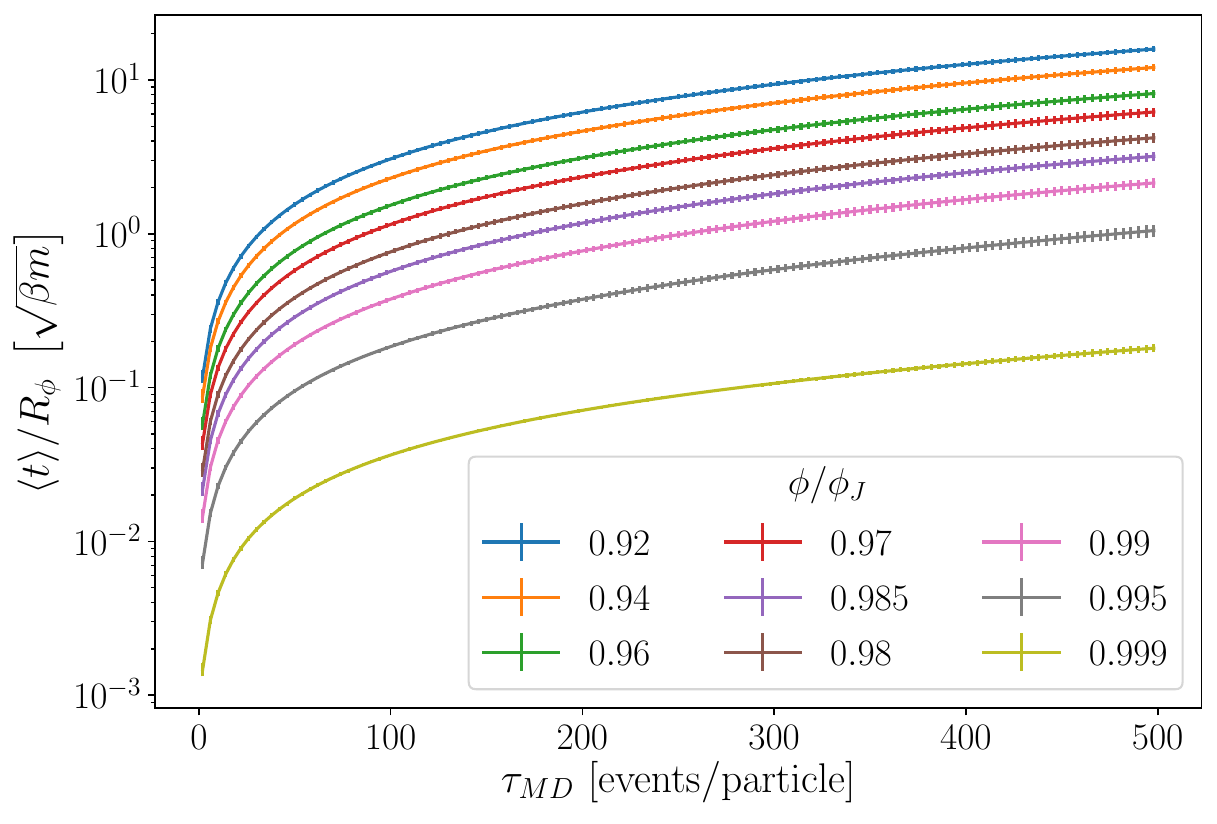}
	\caption{Relation of the (average) global time as MD simulations evolve as a function of the number of collision events, for different packing fractions (different colours). Each marker correspond to an instant at which the particles positions where stored and the solid lines are just a guide to the eye. Errorbars indicate the value of the standard deviation of the values.}
	\label{fig:collisions-vs-t}
\end{figure}

A more important point is to understand why the contact \emph{vectors} ($\vb{n}_{ij}$) are variables more informative about the dynamics near jamming than the contact \emph{forces} ($\vb{F}_{ij}$). As explained in the main text, this feature can be naturally understood in the case of predicting the direction of motion, but this is no longer the case when we consider the particles mobility. While in the main text we already provided some intuition about why we should prefer a description in terms of $\vb{n}_{ij}$ instead of the corresponding force, here we turn to show that even though they are clearly related, their probability distributions (pdf) are strikingly different. This claim is summarized in Fig.~\ref{fig:dist-contacts}, where left panels correspond to distributions involving contact forces while the right ones are the pdf's of the analogous quantity using contact vectors. Firstly, let us consider the quantities in the panels (a) and (b), namely the pdf of dot products between pairs of contact forces and contact vectors, respectively. At a first glance, the distribution associated with the contact forces seems to indicate that the vast majority of contact forces are nearly orthogonal (notice the log-scale in the vertical axis), but when comparing with the result from the contact vectors we see that the likeliest arrangement is the one consisting in three spheres touching each other (thus the peak at $\cos \pi/3=0.5$), while any other formation is roughly equally likely. This in turn implies that the peak at zero evident in the first case is due to the existence of many small forces whose effect is to blur any intrinsic geometrical feature of the configuration's structure.

%

The distinction between these two variables is more evident when analysing the distribution of the sums $\displaystyle \sum_{j<k \in \partial i} \vb{F}_{ij}\cdot \vb{F}_{ik}$ and $\displaystyle \sum_{j<k \in \partial i} \vb{n}_{ij}\cdot \vb{n}_{ik}$, as shown in the lower panels of Fig.~\ref{fig:dist-contacts}. When considering the forces, panel (c) of this figure suggests that not only there is an excess of very small forces, but there are particles whose full set of contact forces is given by forces of minute magnitude.
Even more, the probability of having a different value for this variable decreases exponentially, $p(x) \sim e^{0.22 x}$ , as indicated by the red dashed line. Such a pdf is clearly different from the analogous one obtained using contact vectors, as shown in panel (d). In this latter case, the distribution is mainly concentrated at an intermediate value and then decreases in a Gaussian manner, as depicted by the superimposed black line. Such a behaviour is in line with the observed dynamics, where a small fraction of the particles are either very mobile or very constrained and a significantly larger amount displays similar values of mobility. The comparison of these variables provides additional evidence about the fact that the contact forces do not contain enough information about the dynamics in the vicinity of a jammed configuration.

\begin{figure}[!htb]
	\centering
	\includegraphics[width=0.99\linewidth]{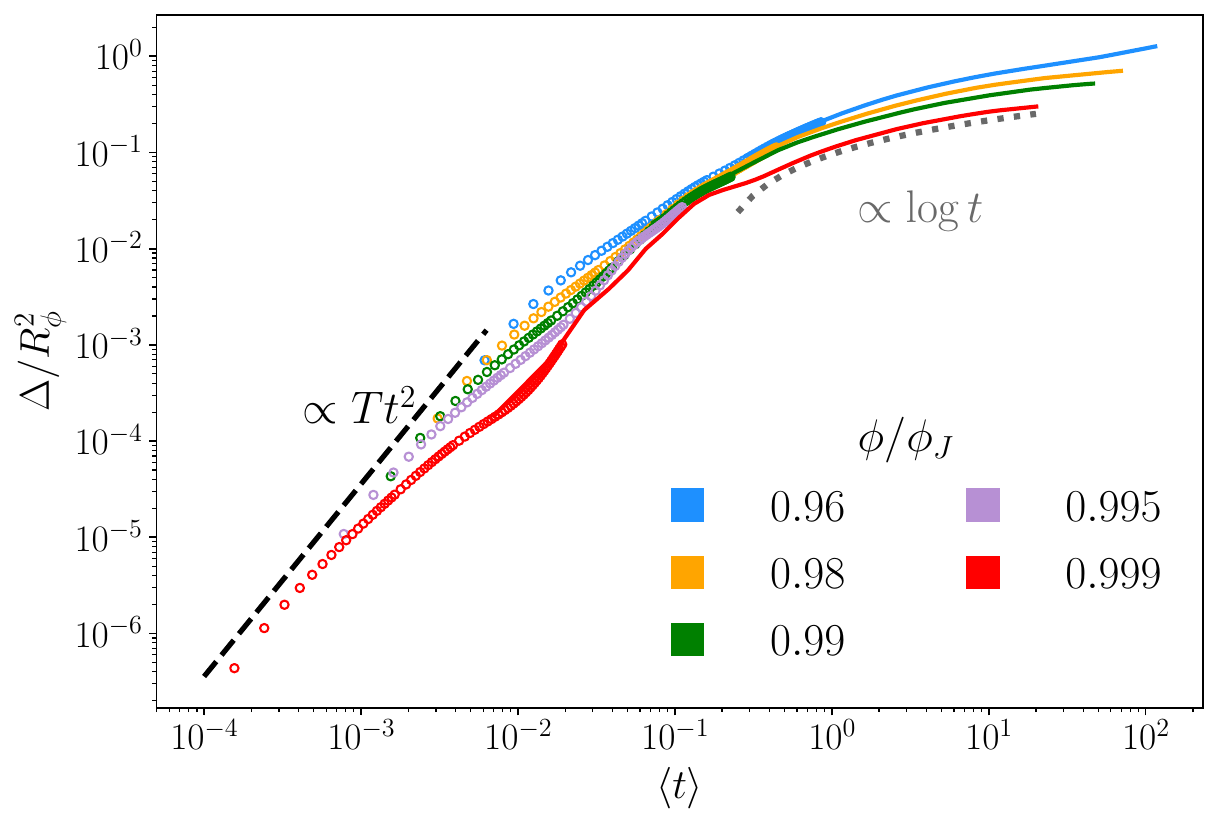}
	\caption{Temporal evolution of the MSD for different packing fractions (different colours). The circular markers show the values of MSD obtained from the main simulations used in this work, while the solid lines correspond to simulations 100 as longer, but for a reduced number of values of $\phi$. Reference lines to compare with well characterize dynamical regimes are also included. Note that all the results presented in the main text correspond to the evolution of the configuration before the logarithmic grow $\Delta$ takes over, which is the regime that so far has received most attention.}
	\label{fig:MD-msd-vs-t}
\end{figure}

\section{Details of MD simulations} \label{sec:supplementary MD}

\begin{figure*}[!htb]
	\centering
	\includegraphics[width=0.85\textwidth]{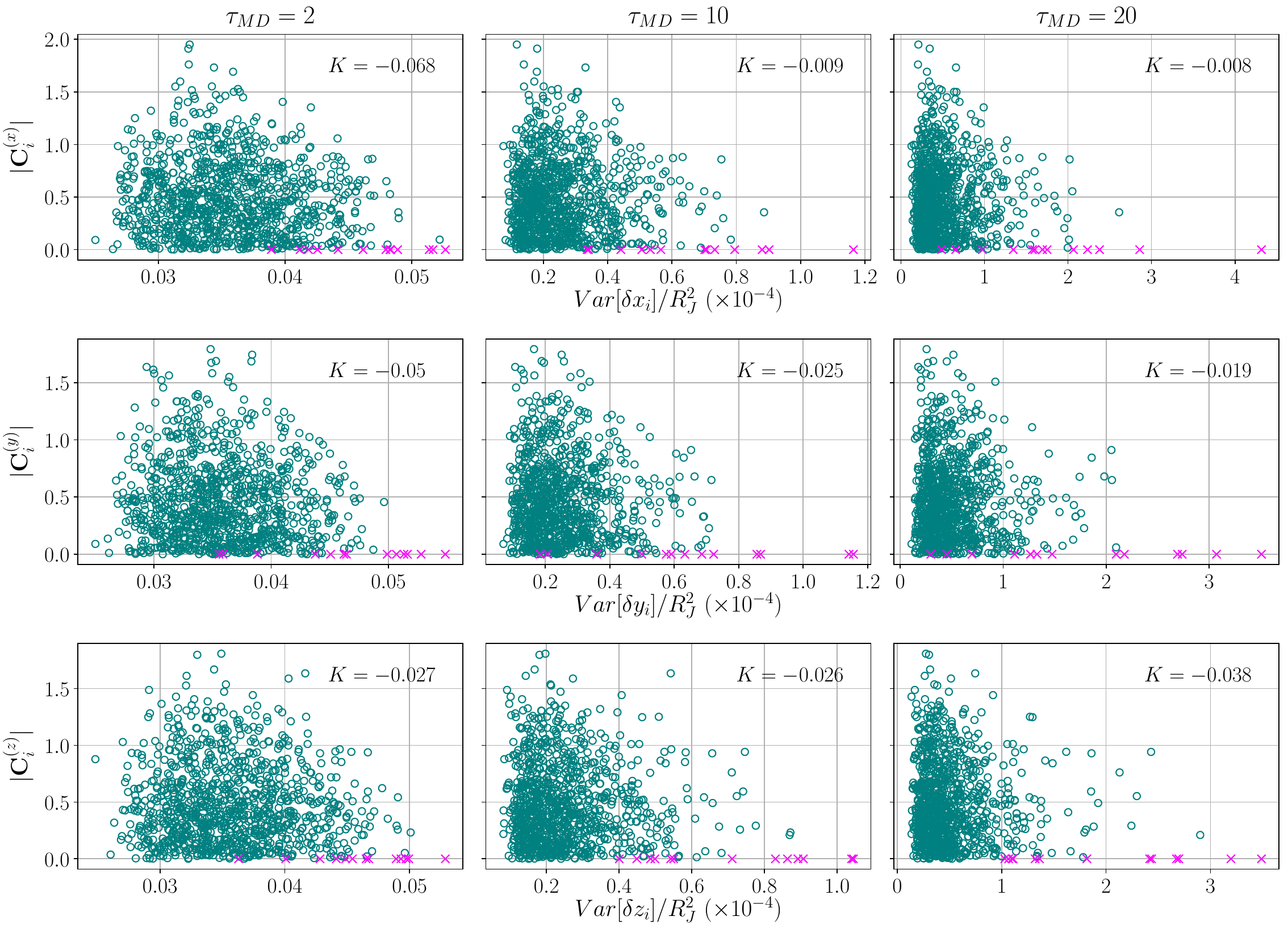}
	\caption{Scatter plot of the components (one in each row) of $Var[\delta \vb{r}_i]$ and the absolute value of the components of $\vb{C}_i$ in the MD simulations at different times (different columns). Rattlers are identified by pink crosses as before.}
	\label{fig:MD-variance-vs-totCV}
\end{figure*}

As explained in the main text, we used event-driven MD simulations to explore the dynamics of hard-sphere systems using the algorithm of Ref.~\cite{md-algorithm}. Because hard-sphere potential renders the dynamics trivial, once the particles have been assigned their initial velocities, we can predict which particles will collide first and at what time. Then the state of the configuration is advanced until that moment and since collisions are assumed to be elastic, the new velocities of the two particles can be computed exactly so the next collision event can also be predicted. In such a way, the evolution of the system is updated each time a collision event occurs and the global time $t$ of the simulation is just the sum of the times passed between each event. Another consequence of using the hard-sphere potential is that the only characteristic length of the system is the particles size, which in turn fixes the time units as $\sqrt{\beta m R^2}$, where $m$ and $R$ are the spheres mass and diameter, respectively. Without loss of generality, we set $m=1$, $\beta=10$, and $R$ is clearly a function of the value of $\phi$ we chose to run the simulations. In Fig.~\ref{fig:collisions-vs-t} we show that for a fixed value of the packing fraction, there is well defined relation between the $t$, averaged over all the trajectories, and the value of $\tau_{MD}$, which measures the number of collisions that have taken place. The very small errorbars suggest that the distribution of times is considerably peaked around $\avg{t}$, which justifies why we used $\avg{t}=t$ in the main text.
%

\begin{figure}[!htb]
	\centering
	\includegraphics[width=0.99\linewidth]{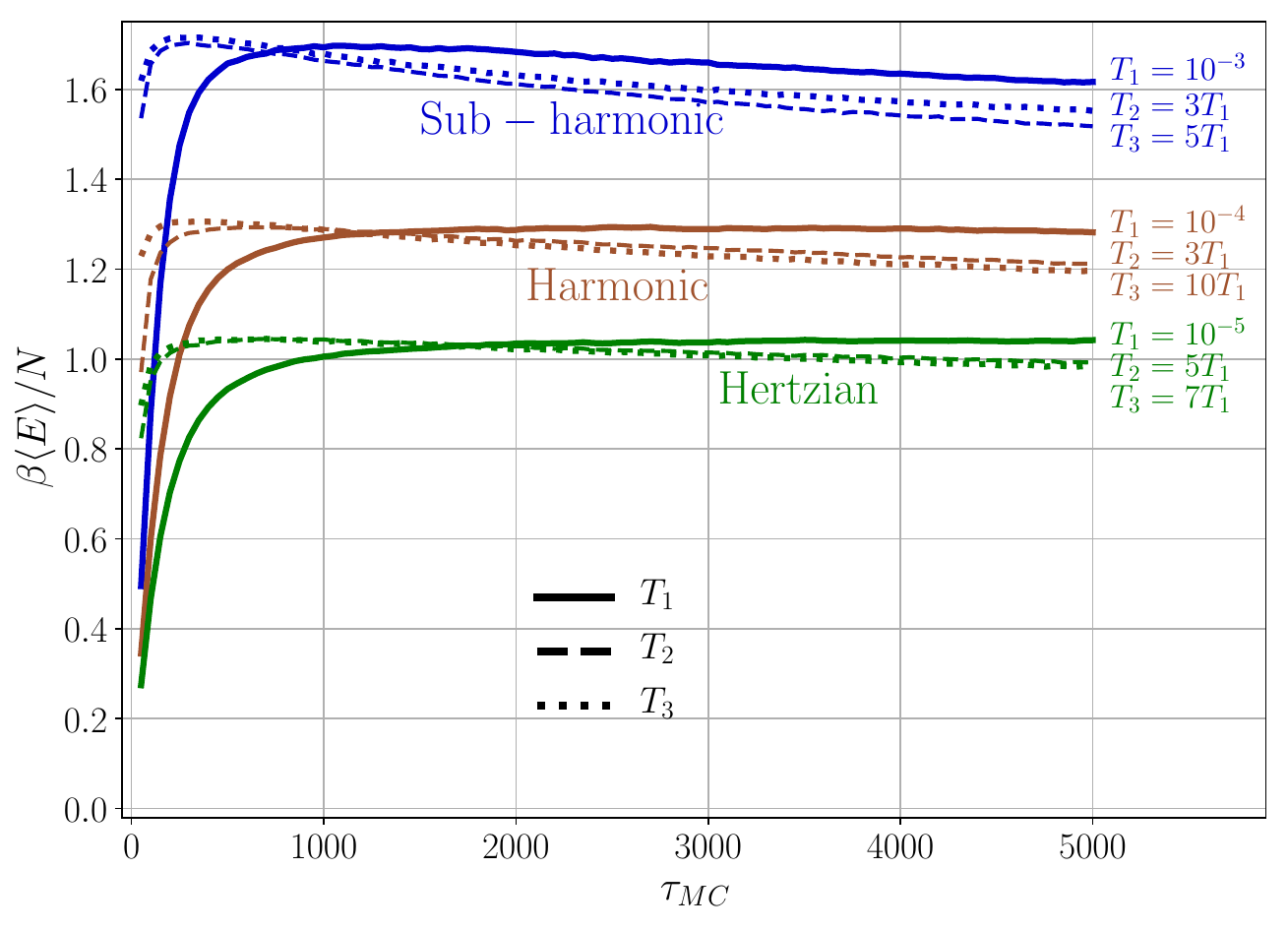}
	\caption{Average energy per particle as a function of the number of MC steps ($\tau_{MC}$) for the three potentials used: $\nu=3/2$ or sub-harmonic (blue), $\nu=2$ or harmonic (brown), and $\nu=5/2$ or Hertzian (green). Different line styles correspond to different temperatures ($T_1<T_2<T_3$), as indicated in the right legends.}
	\label{fig:MC-energy-vs-t}
\end{figure}

To better characterize the dynamical regime studied in this work, we performed additional MD simulations letting the configuration evolve for $100$ times as many collision events, using configurations of packing fractions $\phi/\phi_J = \{0.96, 0.99,0.98, 0.999\}$ and a smaller number of trajectories, namely $M_{long}=1000$. In Fig.~\ref{fig:MD-msd-vs-t} we present the results of the MSD, $\Delta$ as a function of time, for several packing fractions using the data of the main simulations (circular markers) and from the longer ones (solid lines). We also included reference curves to compare with the common dynamical regimes: ballistic (black dashed curve) and logarithmic (grey dotted). It is a well known fact that most of the unique features of glassy systems occur in the latter one, but as this figure shows we are exploring the configuration's evolution well before such a behaviour begins. We should also point out that the ballistic regime is almost entirely lost due to the event-driven nature of the simulations which does not allow us to access the evolution before collisions take place.
%

\begin{figure*}[!htb]
	\centering
	\includegraphics[width=0.85\textwidth]{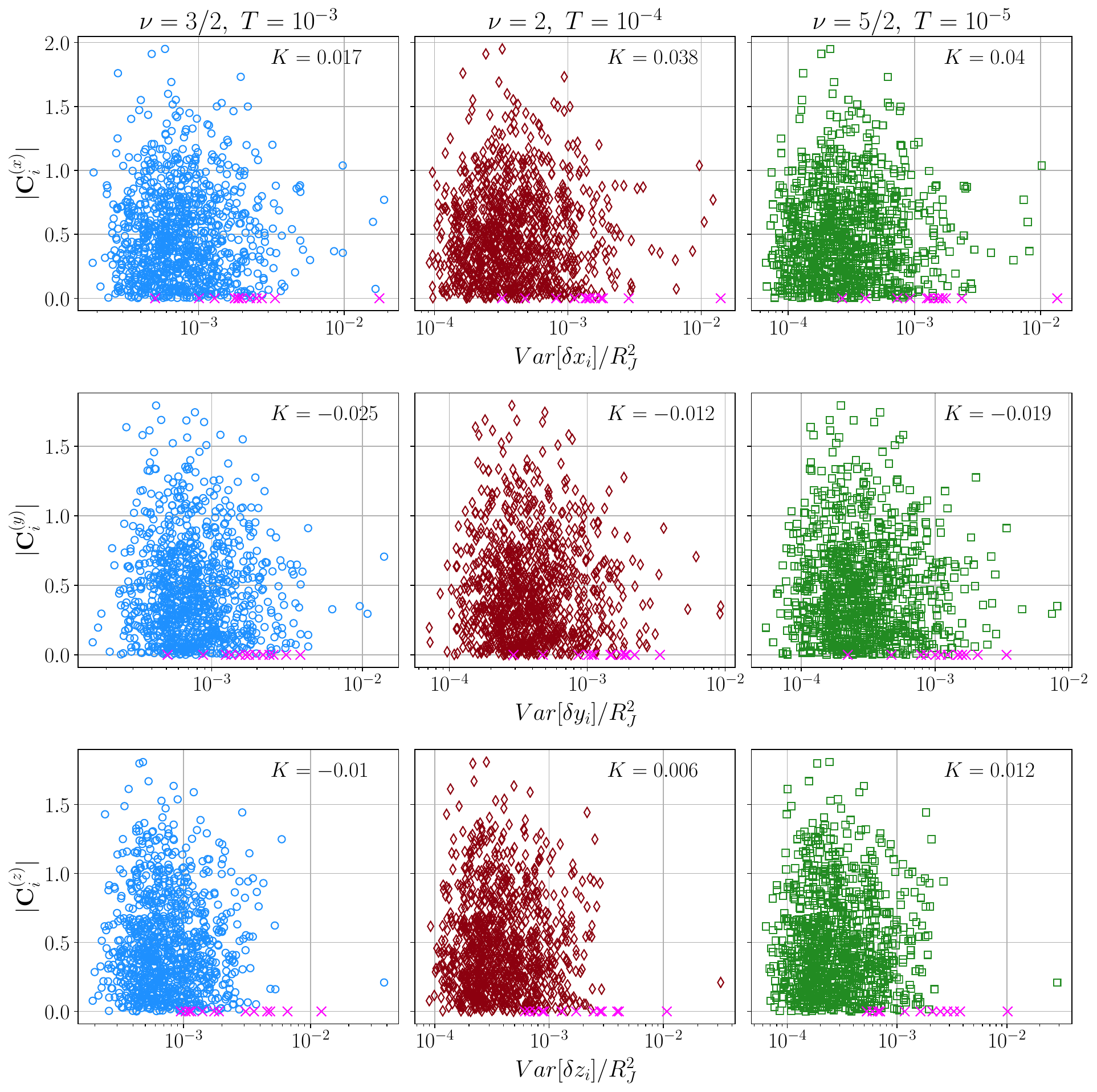}
	\caption{Scatter plot of the components (one in each row) of $Var[\delta \vb{r}_i]$ and the absolute value of the components of $\vb{C}_i$ in the MC simulations for the three different potentials considered in this work (different columns). All the data reported here correspond to $\tau_{MC}=250$ and the value of $T$ used throughout the simulations of each potential are reported above the corresponding column. Different colours are used for the sake of clarity, except for the rattlers data, which as usual is depicted with the pink crosses.}
	\label{fig:MC-variance-vs-totCV}
\end{figure*}

To conclude this section, in Fig.~\ref{fig:MD-variance-vs-totCV} we present the results of a component-wise comparison of $Var[\delta \vb{r}_i]$ with the sum of contact vectors, $\vb{C}_i$, in absolute value. As mentioned in the main text, there is no significant relation between these two quantities and the values of $K$ found in this case are notably small, reflecting the fact that contact vectors only provide information about preferential directions in the particle's motion, but not about the fluctuations around them.
%

\section{Analysis of MC simulations} \label{sec:supplementary MC}

Here we present some details about the MC simulations performed by changing the hard-sphere potential to a soft, contact one, as defined in Eq.~\eqref{def:potential soft sphere} with $\nu=\{3/2, 2, 5/2\}$. In all cases, we kept the spheres' radius fixed at $R_J$ and used the Metropolis-Hastings algorithm to run the dynamics with a constant temperature $T$. In Fig.~\ref{fig:MC-energy-vs-t} we present the temporal evolution of the average energy per particle, scaled by the inverse temperature, $ \beta \avg{E}/N$, for the three interactions potentials used and the different temperatures used for each one. The figure shows that for a fixed values of $\nu$, the energies of the different simulations are comparable between them, once the scaling with $\beta$ is considered. Additionally, the fact that even different interaction potentials yield values of $\beta \avg{E}$ within a small range indicates that the dynamics with different temperatures and potentials are similar and thus belong to the same regime.
%

\begin{figure*}[!htb]
	\centering
	\includegraphics[width=0.85\linewidth]{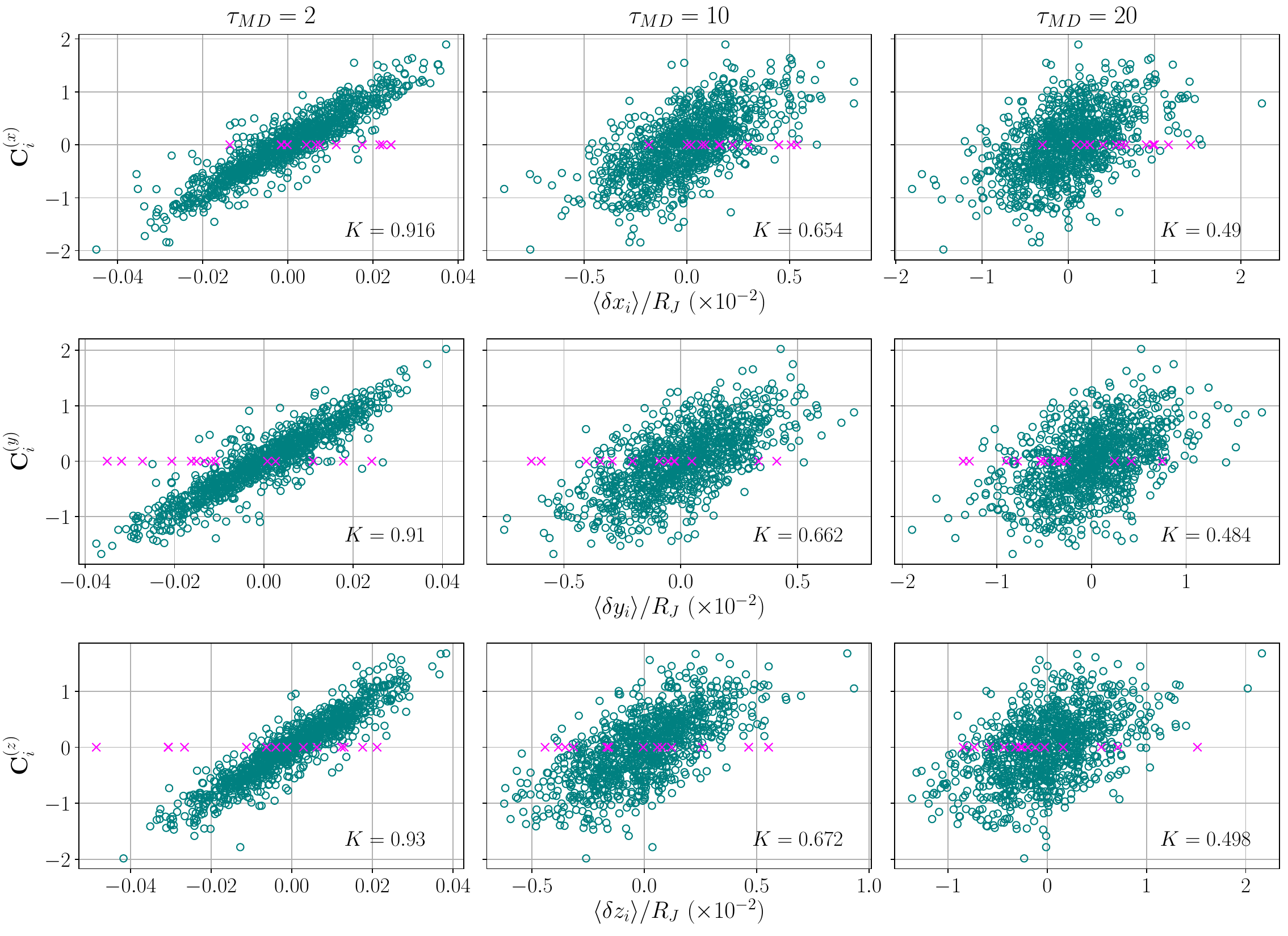}
	\caption{Correlations of mean displacements and $\vb{C}_i$ for a second independent configuration. Note the close resemblance to the results reported in Fig.~\ref{fig:MD-displacements-vs-totCV}. The packing fraction used is $\phi/\phi_J=0.995$.}
	\label{fig:MD-displacements-vs-totCV-2nd-config}
\end{figure*}

Finally, in Fig.~\ref{fig:MC-variance-vs-totCV} we present the analogous of Fig.~\ref{fig:MD-variance-vs-totCV} but using the MC dynamics. In agreement with that previous case, the figure demonstrates that the sum of contact vectors is not a good indicator of the single-particle displacement's variance, even when the interaction potential has been softened. We present the results of all the components (one in each row) and the three interaction types used (one in each column), for a fixed time of $\tau_{MC}=250$ and a single temperature, indicated on top of each column.
%

\section{Reduced analysis on an independent configuration} \label{sec:second-configuration}

\begin{figure}[!htb]
	\centering
	\includegraphics[width=0.99\linewidth]{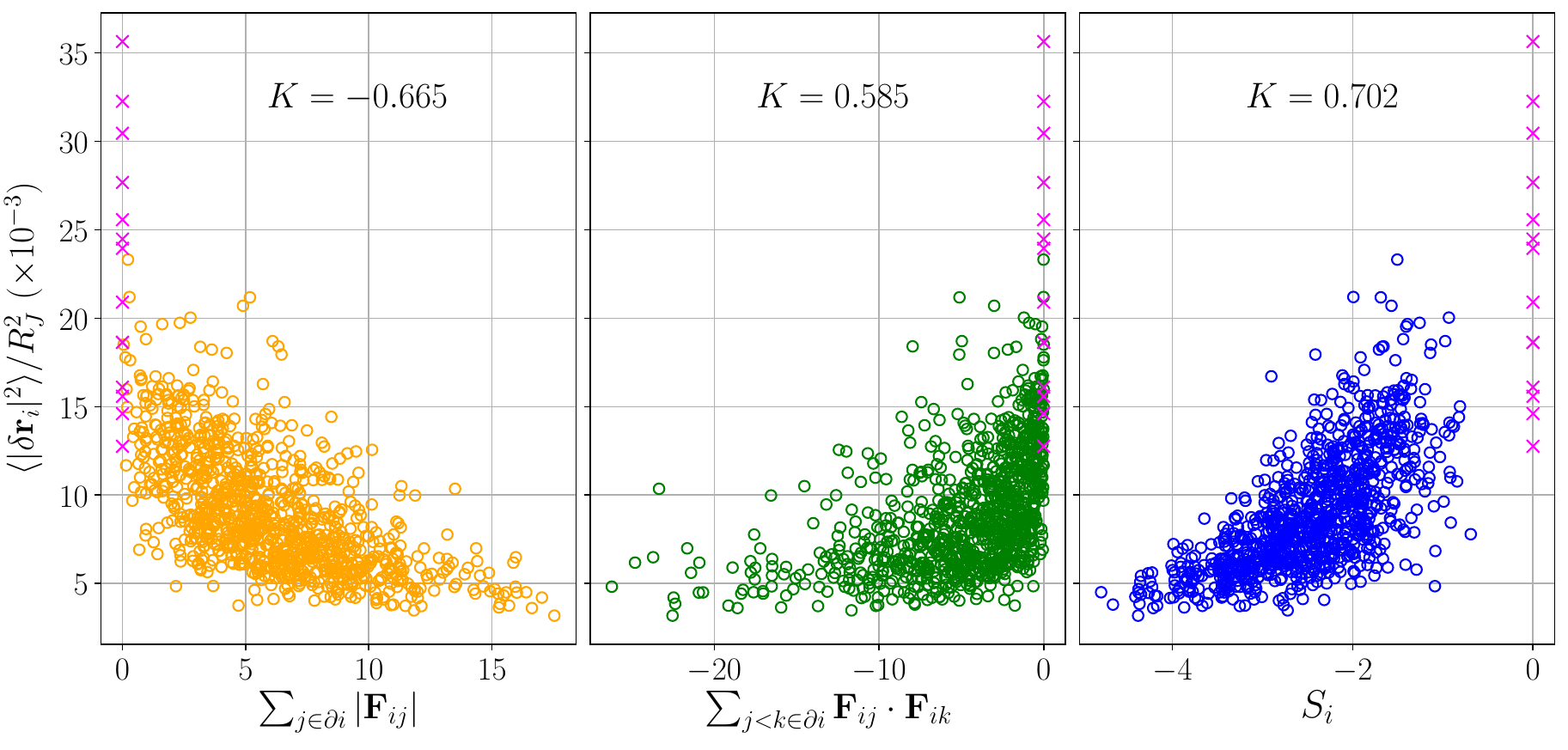}
	\caption{Correlations of particles' mobility and three different structural variables obtained from the network of contact forces. Note that these results reproduce the ones reported in Fig.~\ref{fig:MD-mobility-vs-dotCV}. The packing fraction used is $\phi/\phi_J=0.995$.}
	\label{fig:MD-mobility-vs-dotCV-2nd-config}
\end{figure}

To verify that out methodology is system independent, we performed the same analysis on an independent configuration, also brought to its jamming point via our LP algorithm. As before, we used a configuration of $N=1024$ particles, including $1.4\%$ of rattlers, and whose packing fraction was $\phi_J=0.6350$. We analysed a smaller set of packing fractions using the same MD algorithm and found very similar results for the correlations of both types of structural variables. For instance, in Fig.~\ref{fig:MD-displacements-vs-totCV-2nd-config} the scatter plots showing the correlations between the mean displacement of particles and the corresponding values of $\vb{C}_i$ are shown, analogous to Fig.~\ref{fig:MD-displacements-vs-totCV}. Similarly, the corresponding results where we verify that once again the value of $S_i$ is a better predictor of the particles mobility than the other scalar variables considered is shown in Fig.~\ref{fig:MD-mobility-vs-dotCV-2nd-config}. We thus confirm the choice of $S_i$ as the right structural variable instead of using the ones where the magnitudes of the contact forces are considered.
%

\begin{figure}[!htb]
	\centering
	\includegraphics[width=1.02\linewidth]{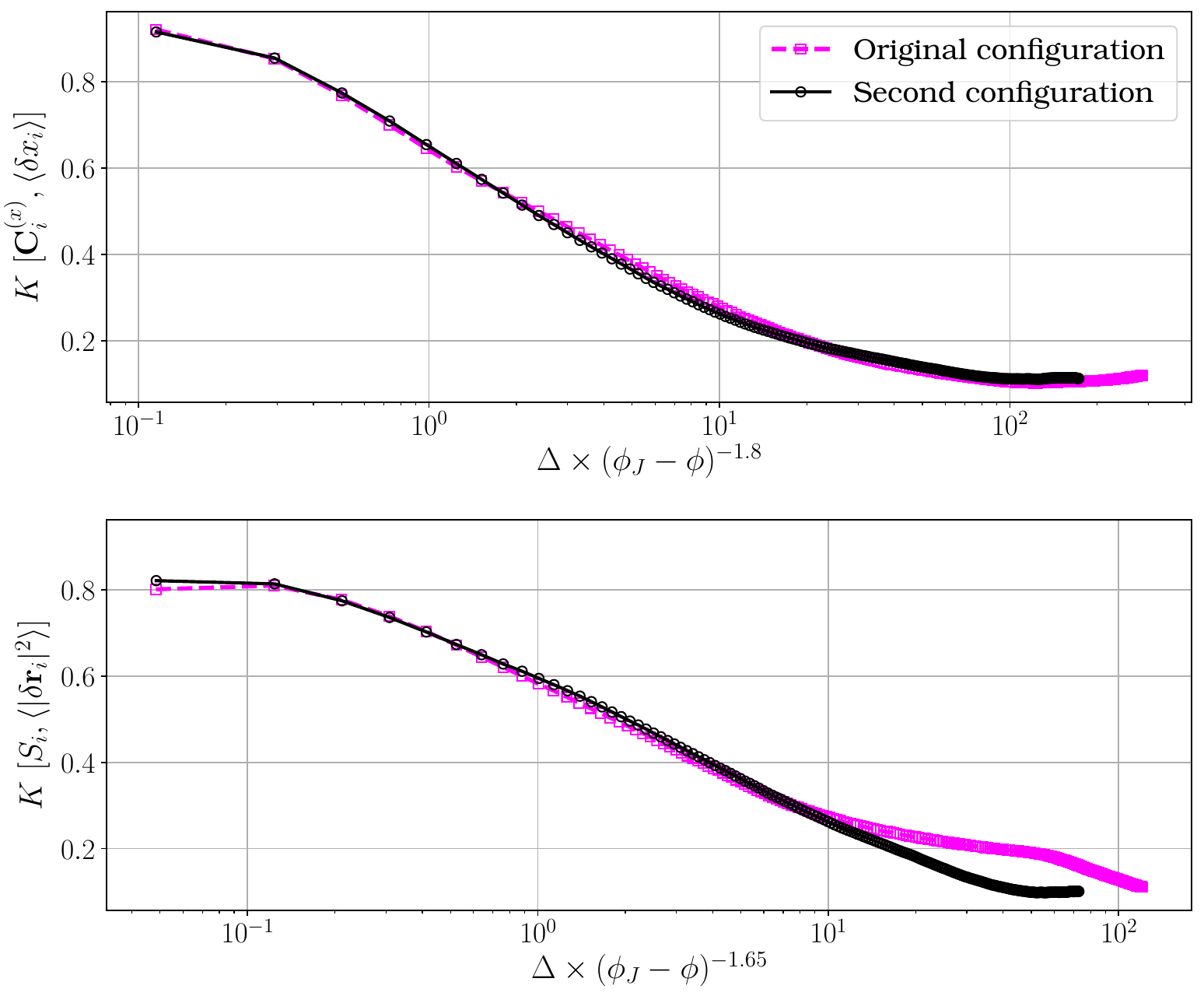}
	\caption{Comparison of the temporal evolution of the statistical correlations between structure and dynamics in two independent configurations.}
	\label{fig:correlations-vs-t-2nd-config}
\end{figure}

Finally, we also tested that the decorrelation from the initial configuration follows the same universal behaviour as the ones reported in Fig.~\ref{fig:correlations-vs-t}. To do so, we computed the values of $K[\avg{\delta \vb{r}_i}, \vb{C}_i]$ and $K[\avg{|\delta \vb{r}_i|^2}, S_i]$ as they evolved in time, and compared them with the ones of the original configuration at the corresponding packing fraction. The comparison is shown in Fig.~\ref{fig:correlations-vs-t-2nd-config}; the fact that the two curves resemble each other very closely shows that our approach is robust enough to be applied to any jammed configuration of spherical particles.

\renewcommand\refname{References}

\bibliography{all_refs} 
\bibliographystyle{rsc} 

\end{document}